\documentclass[amssymb,prb,twocolumn,showpacs,superscriptaddress]{revtex4}
\usepackage{graphicx}
\usepackage{epsfig}
\usepackage{amsmath}
\newfont{\bl}{cmbxsl10 scaled\magstep1}
\begin{document}

\title{Paramagnetic GaN:Fe and ferromagnetic (Ga,Fe)N -- relation between structural, electronic, and magnetic properties}

\author{Alberta Bonanni}
\affiliation{Institut f\"ur Halbleiter- und Festk\"orperphysik, Johannes Kepler University, Altenbergerstr. 69, A-4040 Linz, Austria}
\author{Micha{\l} Kiecana}
\affiliation{Institute of Physics,
Polish Academy of Sciences, al.~Lotnik\'ow 32/46, PL 02-668 Warszawa, Poland}
\author{Clemens Simbrunner}
\affiliation{Institut f\"ur Halbleiter- und Festk\"orperphysik, Johannes Kepler University, Altenbergerstr. 69, A-4040 Linz, Austria}
\author{Tian Li}
\affiliation{Institut f\"ur Halbleiter- und Festk\"orperphysik, Johannes Kepler University, Altenbergerstr. 69, A-4040 Linz, Austria}
\author{Maciej Sawicki}
\affiliation{Institute of Physics, Polish Academy of Sciences, al.~Lotnik\'ow 32/46, PL 02-668 Warszawa, Poland}
\author{Matthias Wegscheider}
\affiliation{Institut f\"ur Halbleiter- und Festk\"orperphysik, Johannes Kepler University, Altenbergerstr. 69, A-4040 Linz, Austria}
\author{Martin Quast}
\affiliation{Institut f\"ur Halbleiter- und Festk\"orperphysik, Johannes Kepler University, Altenbergerstr. 69, A-4040 Linz, Austria}
\author{Hanka Przybyli\'nska}
\affiliation{Institut f\"ur Halbleiter- und Festk\"orperphysik, Johannes Kepler University, Altenbergerstr. 69, A-4040 Linz, Austria}
\affiliation{Institute of Physics,
Polish Academy of Sciences, al.~Lotnik\'ow 32/46, PL 02-668 Warszawa, Poland}
\author{Andrea Navarro-Quezada}
\affiliation{Institut f\"ur Halbleiter- und Festk\"orperphysik, Johannes Kepler University, Altenbergerstr. 69, A-4040 Linz, Austria}
\author{Rafa{\l} Jakie{\l}a}
\affiliation{Institute of Physics,
Polish Academy of Sciences, al.~Lotnik\'ow 32/46, PL 02-668 Warszawa, Poland}
\author{Agnieszka Wolos}
\affiliation{Institut f\"ur Halbleiter- und Festk\"orperphysik, Johannes Kepler University, Altenbergerstr. 69, A-4040 Linz, Austria}
\author{Wolfgang Jantsch}
\affiliation{Institut f\"ur Halbleiter- und Festk\"orperphysik, Johannes Kepler University, Altenbergerstr. 69, A-4040 Linz, Austria}
\author{Tomasz Dietl}
\affiliation{Institute of Physics, Polish Academy of Sciences, al.~Lotnik\'ow 32/46, PL 02-668 Warszawa, Poland}
\affiliation{ERATO Semiconductor Spintronics Project, Japan Science and Technology Agency, al.~Lotnik\'ow 32/46, PL 02-668 Warszawa, Poland}
\affiliation{Institute of Theoretical Physics, Warsaw University, ul.~Ho\.za 69, PL 00-681 Warszawa, Poland}

\date{\today}
\begin{abstract}
We report on the metalorganic chemical vapor deposition (MOCVD) of
GaN:Fe and (Ga,Fe)N layers on c-sapphire substrates and their thorough
characterization {\em via} high-resolution x-ray diffraction
(HRXRD), transmission electron microscopy (TEM), spatially-resolved
energy dispersive X-ray spectroscopy (EDS), secondary-ion mass
spectroscopy (SIMS), photoluminescence (PL), Hall-effect,
electron-paramagnetic resonance (EPR), and magnetometry employing a
superconducting quantum interference device (SQUID). A combination
of TEM and EDS reveals the presence of coherent nanocrystals
presumably Fe$_x$N with the composition and lattice parameter
imposed by the host. From both TEM and SIMS studies, it is stated
that the density of nanocrystals and, thus the Fe concentration
increases towards the surface. According to Hall effect
measurements, electrons from residual donors are trapped by mid-gap
Fe acceptor states in the limit of low iron content $x
\lesssim 0.4$\%, indicating that the concentration of Fe$^{2+}$ ions
increases at the expense of Fe ions in the 3+ charge state. This
effect is witnessed by photoluminescence (PL) measurements as
changes in the intensity of the Fe$^{3+}$-related intra-ionic
transition, which can be controlled by co-doping with Si donors and
Mg acceptors. In this regime, EPR of Fe$^{3+}$ ions and Curie-like
magnetic susceptibility are observed. As a result of the spin-orbit
interaction, Fe$^{2+}$ does not produce any EPR response. However,
the presence of Fe ions in the 2+ charge state may account for a
temperature-independent Van Vleck-type paramagnetic signal that we
observe by SQUID magnetometry. Surprisingly, at higher Fe
concentrations, the electron density is found to increase
substantially with the Fe content. The co-existence of electrons in
the conduction band and Fe in the 3+ charge state is linked to the
gradient in the Fe concentration. In layers with iron content $x
\gtrsim 0.4$\% the presence of ferromagnetic signatures, such as
magnetization hysteresis and spontaneous magnetization, have been
detected. A set of precautions has been undertaken in order to rule
out possible sources of spurious ferromagnetic contributions. Under
these conditions, a ferromagnetic-like response is shown to arise
from the (Ga,Fe)N epilayers, it increases with the iron
concentration, it persists up to room temperature, and it is
anisotropic -- {\em i.e.}, the saturation value of the magnetization
is higher for in-plane magnetic field. We link the presence of
ferromagnetic signatures to the formation of Fe-rich nanocrystals,
as evidenced by TEM and EDS studies. This interpretation is
supported by magnetization measurements after cooling in- and
without an external magnetic field, pointing to superparamagnetic
properties of the system. It is argued that the high temperature
ferromagnetic response due to spinodal decomposition into regions
with small and large concentration of the magnetic component is a
generic property of diluted magnetic semiconductors and diluted
magnetic oxides showing high apparent Curie temperature.
\end{abstract}

\maketitle

\section{Introduction}
In recent years, it has become more and more clear that wide band-gap
semiconductors and oxides doped with transition metals (TM) constitute a
new class of material systems exhibiting magnetic properties whose origin
and methods of control are still not
understood.\cite{Dietl:2003_z,Pearton:2003_a,Prellier:2003_a,Dietl:2005_c,Liu:2005_a}
To this category belongs certainly (Ga,Fe)N, which is the subject of the
present work. While extensive studies have been conducted on
(Ga,Mn)N\cite{Pearton:2003_a,Graf:2003_a,Dietl:2004_z,Liu:2005_a} as
promising workbench for future applications in
spintronics,\cite{Matsukura:2002_c,Wolf:2001_a} only little is known about
the (Ga,Fe)N materials system.

The magnetic properties of nominally undoped and $p$-doped GaN
implanted with Fe ions were reported, showing a ferromagnetic
response evidenced by magnetization hysteresis loops persisting,
depending on the provided dose,\cite{Pearton:2002_a} up to room
temperature (RT).\cite{Abernathy:2002_a,Shon:2004_a} Furthermore,
the emission channeling technique applied to Fe-implanted (Fe dose
up to $10^{19}$~cm$^{-3}$) GaN samples confirmed the presence of a
high percentage of TM ions (up to 80\%) occupying substitutional Ga
sites of the host crystal.\cite{Wahl:2001_a}  Moreover, GaN films
doped with Fe, with concentrations up to
$~3$~$\times$~10$^{19}$~cm$^{-3}$ were grown by molecular beam
epitaxy (MBE) at substrate temperatures $T_{\mathrm{S}}$ ranging
from 20 to $380^{\circ}$C directly on sapphire (0001) and
ferromagnetic behavior with a Curie temperature
$T_{\mathrm{C}}$~=~100~K has been observed only in the samples grown
at $~400^{\circ}$C.\cite{Akinaga:2000_a} Films of GaN:Fe (Fe density
up to $6$~$\times$~10$^{21}$~cm$^{-3}$) fabricated by means of MBE
at $T_{\mathrm{S}} ~=~$500~--~800$^{\circ}$C showed a
superparamagnetic behavior\cite{Kuwabara:2001_a,Kuwabara:2001_b}
assigned to Ga-Fe and/or Fe-N inclusions. Extended x-ray absorption
fine structure (EXAFS) analysis suggests that the decrease of
$T_{\mathrm{C}}$ leads to a structural transition from the wurtzite
(wz) to the zinc-blende (zb) reconstruction, and this transition may
be related to the origin of ferromagnetism in Fe-doped GaN
films.\cite{Ofuchi:2001_a} The metalorganic chemical vapor
deposition (MOCVD) of GaN:Fe has been previously reported, with a
focus on the actual Fe content in the layers and its effect onto the
carrier concentration.\cite{Mishra:2003_a} More recent preliminary
works by
others\cite{Kane:2006_a} and us\cite{Przybylinska:2006_a,Bonanni:2006_a} indicate that MOCVD grown (Ga,Fe)N shows
ferromagnetic-like characteristics to above room temperature.
Moreover, diluted magnetic semiconductors, and (Ga,Fe)N in
particular, have become model systems to test various
implementations of the density functional theory to disordered
strongly correlated
systems.\cite{Sato:2002_a,Mirbt:2002_a,Sanyal:2003_a,Cui:2006_a}

Despite the mentioned previous reports, a thorough characterization
of this novel materials system, with special attention to the
mechanisms of the magnetic response, is still needed. In order to
predict the feasibility of e.~g. carrier-mediated spin coupling, it
is crucial to know the electronic and spin structure of the TM
centers, which depends on the actual charge state and is strongly
influenced by the host crystal. Concerning the position of the deep
acceptor-like level Fe$^{3+}$/Fe$^{2+}$ in the GaN gap, it was
estimated from photoluminescence (PL) excitation spectroscopy
studies to be 2.5~--~2.6~eV above the GaN valence band edge
$E_{v}$.\cite{Maier:1994_a,Baur:1994_a} However, another group
suggests that this level may lie much closer to the conduction band
at ($E_{v}$~+~2.863)~$\pm$ ~0.005~eV.\cite{Malguth:2006_b}

The present work is devoted to a comprehensive study of the GaN:Fe
(below the solubility limit of Fe into GaN) and (Ga,Fe)N material
systems, beginning with a careful on-line control of the growth
process and proceeding with a possibly thorough investigation of the
structural, electrical, optical, and magnetic properties in order to
shed new light into the mechanisms responsible for the paramagnetic
and high-temperature ferromagnetic response of these novel systems.
Particular attention is paid to avoid contamination of the
superconducting quantum interference device (SQUID) signal by
spurious effects and, thus, to reliably determine the magnetic
properties of (Ga,Fe)N. By combining various characterization
techniques, we establish experimentally the relation between
structural, electronic, and magnetic properties of the studied
material.  We show how doping over the solubility limit affects the
sample morphology and magnetic ion distribution, which in turn
affect in a dramatic way both electronic and magnetic properties of
the system. The interpretation of our results in terms of spinodal
decomposition and non-uniform distribution of magnetic nanocrystals
along the growth direction can be applied to explain
ferromagnetic-like properties persisting up to high temperatures in
a broad class of diluted magnetic semiconductors and diluted
magnetic oxides.

Our paper is organized as follows. In the next section we
describe the experimental methods employed to characterize 
{\em in-situ} and {\em ex-situ} the structural, optical, transport and magnetic
properties of the MOCVD structures under investigation. Section III contains a summary of sample
parameters, including Fe concentrations, as determined by various
experimental methods. In Sec.~IV we present experimental results
starting from findings obtained from transmission electron
microscopy (TEM) and spatially-resolved energy dispersive X-ray
spectroscopy (EDS). These measurements reveal a highly non-uniform
distribution of the magnetic ions at Fe concentrations surpassing
the solubility limit. We then discuss how the presence of magnetic
nanocrystals and the non-uniformity in their distribution modify the
electrical, optical, magnetic resonance, and magnetic
characteristics of the system. Conclusions and outlook stemming from
our work are summarized in Sec.~V. Finally, in Appendix, we  discuss
in detail the procedure employed to determine reliably the magnetic
response of (Ga,Fe)N by SQUID magnetometry.

\section{Experimental}
\subsection{Growth procedure}
The studied epilayers have been fabricated in an AIXTRON 200RF
horizontal-tube MOCVD reactor. All structures have been deposited on
c-plane sapphire substrates according to a well established growth
procedure\cite{Bonanni:2003_a} involving TMGa, NH$_3$, Cp$_2$Mg,
SiH$_4$, and Cp$_2$Fe as precursors for respectively Ga, N, Mg, Si,
and Fe, with H$_2$ as a carrier gas. Upon the nitridation of the
substrate in the reactor, the deposition of a low-temperature
(540~$^{\circ}$C) GaN nucleation layer (LT-NL), its annealing under
NH$_3$ and the growth of a 1~$\mu$m thick device-quality GaN layer
at 1050$^{\circ}$C, as a first step, several series of 500~nm thick
GaN:Fe at different substrate temperatures (ranging from 750 to
1050$^{\circ}$C) and different Cp$_2$Fe flow rates [between 50 and
400~standard cubic centimeters per minute (sccm)] have been
fabricated. The nominal Fe content in subsequently grown samples has
been alternatively switched from low to high and {\em vice-versa},
in order to minimize long term memory effects due to the presence of
residual Fe in the reactor.\cite{Heikmann:2002_a} The samples have
been continuously rotated during growth to promote the deposition
homogeneity.

\begin{figure}[htbp]
    \centering
        \includegraphics[width=8.6cm]{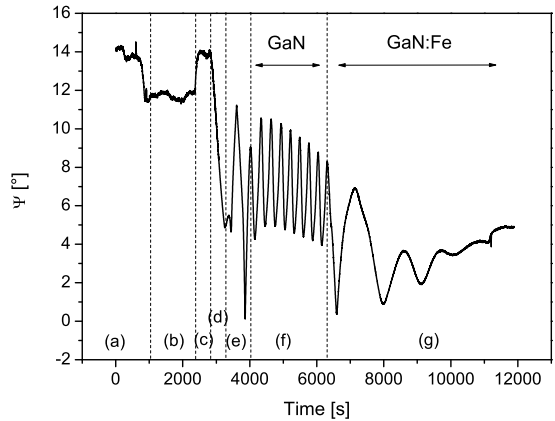}
    \caption{{\em In-situ} ellipsometry: ellipsometric angle $\Psi$ at 2~eV as a function
    of time for a standard deposition sequence. Particular regions relative to the growth
    process and marked (a) to (g) are discussed in the main text.}
    \label{fig:Elli_Psi_def_1}
\end{figure}

\subsection{In-situ monitoring}
\subsubsection{Spectroscopic ellipsometry and laser interferometry}
The employed MOCVD reactor is equipped with an {\em in-situ} Isa
Jobin Yvon ellipsometer which allows both spectroscopic (optical
response -- in terms of the ellipsometric angles $\Psi$ and $\Delta$
-- as a function of the photon energy) and kinetic (optical response
as a function of time) on-line measurements over the energy range
between 1.5 and 5.5~eV for process optimization.\cite{Peters:2000_a}
Figure~\ref{fig:Elli_Psi_def_1} shows the ellipsometric angle $\Psi$
at 2~eV as a function of time for the standard growth process
employed to fabricate the samples under investigation: (a) heating
of the sapphire substrate from RT to 1200$^{\circ}$C; (b) desorption
under H$_{2}$ flow to stabilize the surface; (c) cooling and
nitridation of the substrate; (d) deposition of the LT-NL; (e)
annealing of the NL; (f) growth of the GaN buffer layer; (g) growth
of the Fe doped GaN layer. Spectroscopic ellipsometry  measurements
carried out in real time allow the on-line control over the
growth-rate and the composition of the growing layer and provide
additional qualitative information on the surface roughness.
Moreover, simultaneously to ellipsometry, standard on-line dynamic
optical reflectivity\cite{Balmer:2001_a} at 2~eV has been routinely
performed, yielding analogous information on the growing layer as
ellipsometry, but being more sensitive to wobbling (unavoidable
during the MOCVD process) and temperature
effects.\cite{Bonanni:2005_a}

\subsubsection{In-situ X-ray diffraction}
Furthermore, the growth process has been routinely monitored {\em
in-situ} by x-ray diffraction. In addition to the optical windows
necessary for the on-line ellipsometry characterization, the MOCVD
reactor has been implemented with two Be windows transparent to the
radiation energy provided by a PANalytical Cu x-ray tube. A
Johansson monochromator focuses the beam on the growing crystal in a
geometry suitable for the observation of the (11$\bar2$4) reflection
of wz-GaN. The diffracted beam is collected by a multichannel
detector PANalytical X'Celerator with an angular resolution of
0.013$^{\circ}$ per pixel in this geometry. This diffraction set-up allows us to
perform kinetic measurements yielding information on the layer
composition, thickness and crystalline quality even on rotating
samples, thanks to a recently developed wobbling compensation
algorithm.\cite{Simbrunner:2006_a,Simbrunner:2006_b} Moreover, real
time reciprocal space mapping can be carried out in a wide
temperature range (20~--~1050$^\circ$C), though limited to a static
sample configuration, in order to eliminate wobbling
effects.\cite{Simbrunner:2006_c}

\begin{figure}[htbp]
    \centering
        \includegraphics[width=8.6cm]{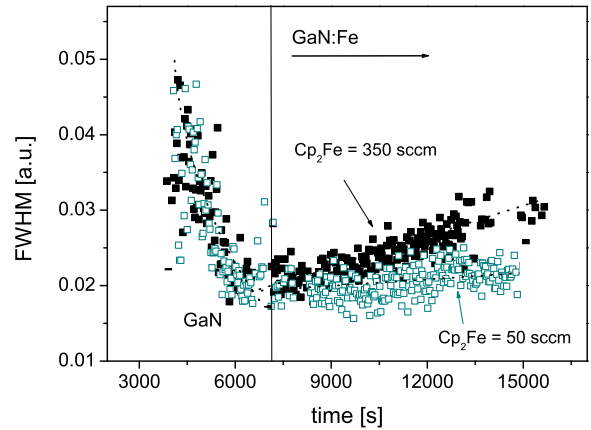}
    \caption{{\emph{In-situ}} XRD kinetic measurement: FWHM of the (11$\bar2$4)
    GaN reflection vs.~deposition time for two samples grown with different Cp$_2$Fe
    flow rates. The evolution of the FWHM during the growth of the GaN buffer is
    reproducibly independent of the sample, whereas the broadening during the
    deposition of the Fe-containing layer is a function of the magnetic ions content.}
    \label{fig:FWHM_Kinetic}
\end{figure}

Kinetic {\emph{in-situ}} XRD measurements were routinely carried out
during growth. A full-width-at-the-half maximum (FWHM) analysis of the GaN peak in the
{\emph{in-situ}} spectra as a function of the growth time, yields
information on the crystal quality of the layers. In
Fig.~\ref{fig:FWHM_Kinetic} a clear decrease of the FWHM during the
deposition of the GaN buffer due to the growing thickness of high
crystalline quality GaN is detectable and was proven to be
quantitatively reproducible on a full series of
samples.\cite{Simbrunner:2006_a} The vertical line demarcates the
time corresponding to the opening of the Cp$_{2}$Fe source and
indicates the onset of broadening of the GaN  peak due to the
incorporation of Fe ions. A clear dependence of the FWHM on the
layer thickness and on the iron precursor flow rate is evident and comparable
to the values obtained from \emph{ex-situ} post-growth XRD analysis
reported in Table I.

\begin{figure}[htbp]
    \centering
        \includegraphics[width=8.6cm]{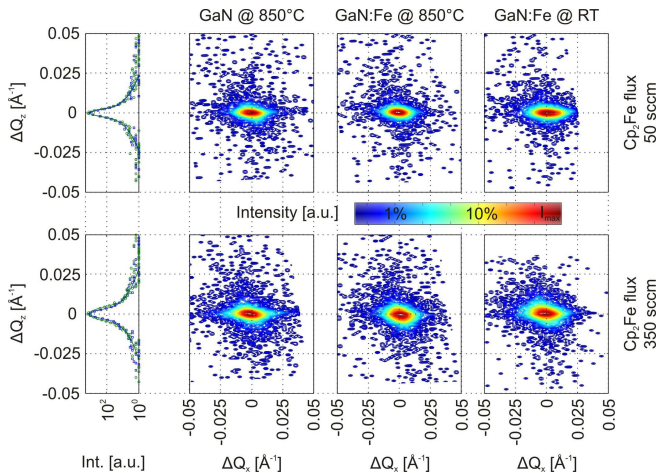}
    \caption{{\emph{In-situ}} XRD space maps in the (11$\bar2$4)
    reflection of GaN and broadening in the inverse of the lattice
    parameter along the growth direction (left panels) for two
    values of Cp$_2$Fe flow rates, 50 and 350 sccm (upper and lower panel, respectively).}
    \label{fig:Maps}
\end{figure}

A more detailed information on the structural properties of the
layers, can be obtained by acquiring {\emph{in-situ}} reciprocal
space maps. The diffraction geometry for the observation of the
(11$\bar2$4) reflection has been chosen in such a way that the
sensitivity to the lattice parameter Q$_{\mathrm{z}}$ perpendicular
to the (0001) surface could be optimized. After the deposition of
the GaN buffer the substrate rotation was interrupted at the proper
azimuth for the acquisition of reciprocal space maps.
Figure~\ref{fig:Maps} presents the space map for the GaN buffer
layer (GaN @ 850$^{\circ}$C) qualitatively reproducible for all
investigated samples. The procedure was repeated just after the
deposition of the Fe-doped layer (GaN:Fe @ 850$^{\circ}$C) and after
cooling the sample to room temperature (GaN:Fe @ RT) upon completion
of the growth process. The reciprocal maps in Fig.~\ref{fig:Maps}
evidence the variation of the lattice parameter along the growth
direction, $\Delta$Q$_{\mathrm{z}}$, obtained from the following
relation,\cite{Simbrunner:2006_a}
\begin{equation}
\Delta Q_{\mathrm{z}} = \frac{2\pi}{\lambda}\cdot
    \cos(\epsilon_{\mathrm{GaN}})\cdot\Delta\epsilon;
\end{equation}
where $\lambda$ is the wavelength of the incident radiation,
$\epsilon_{\mathrm{GaN}}$ gives the diffraction angle for relaxed
GaN and $\Delta\epsilon$ is the measured relative peak position. As
confirmed by {\emph{in-situ}} kinetic measurements and post growth
analysis, the broadening is a function of the iron flow rate.

\subsection{{\sl Ex-situ} characterization methods}
\subsubsection{High-resolution X-ray diffraction}
High-resolution X-ray diffraction (HRXRD) rocking curves of the
GaN~(0002) reflex, give for all the samples, including the (Ga,Fe)N
structures, a FWHM of
200~--~320~arcsec, comparable with state-of-the-art device-quality
gallium nitride material.\cite{Bonanni:2003_b} The broadening of the
GaN(0002) reflex is found to depend both on the iron doping level
and on the growth temperature. The minimum FWHM is obtained for the
lowest Cp$_2$Fe flow rate, 50~sccm, and the highest growth
temperature, 950$^{\circ}$C. Despite the importance of HRXRD for the
routine structural characterization of the layers, the inadequacy of
standard X-ray diffraction methods for nanoscale investigations
(detection of the presence of non-uniform magnetic ion distribution)
is already widely accepted.\cite{Liu:2005_a} We also note that,
according to atomic force microscopy (AFM), the surface roughness
increases with the  Fe concentration, an effect that may affect some
quantitative considerations. The change in sample morphology
indicates the onset of a three-dimensional (3D) granular-like growth
mode characterized by grain height differences exceeding the layer
nominal thickness at the highest Fe concentration.
\subsubsection{Secondary-ion mass spectroscopy}
\label{sec:SIMS} The total Fe concentration in the layers has been
determined via secondary-ion mass spectroscopy (SIMS) and calibrated
with an undoped GaN sample implanted with $10^{16}$~cm$^{-2}$
$^{56}$Fe at 270~keV as reference. The total Fe concentration in the
layers is found to increase with the nominal Cp$_2$Fe flow rate. We
also detect an increase in the oxygen concentration which reaches
10$^{20}$ cm$^{-3}$ at the highest Cp$_2$Fe flow rates, an effect
that may result from a larger surface area due to the
above-mentioned onset of granular growth. We underline that the
granular morphology may render the quantitative determination of Fe
and O contents and profiles not reliable in the case of the highest
Fe concentration.
\subsubsection{Electron-paramagnetic resonance}
Electron paramagnetic resonance (EPR) data have been acquired with a BRUKER ELEXSYS E-580 spectrometer,
operating at a frequency of 9.48 GHz (X-band). The spectrometer is
equipped with a continuous-flow Oxford cryostat, enabling measurements in the
temperature range 2~--~300~K.
\subsubsection{Photoluminescence}
The photoluminescence (PL) measurements in ultraviolet (UV) regime have
been carried out in a set-up consisting of a Helium Cadmium (HeCd)
laser with a wavelength of 325~nm (3.82~eV) and an excitation power
of 40~mW, a Jobyn Yvon monochromator with a focal length of 550~mm
and a Jobyn Yvon 1024~$\times$~512 pixel liquid nitrogen-cooled
Charge Coupled Device (CCD) camera for detection. For the
measurements in the infrared (IR) region the 351.1~nm (3.53~eV) line
of a Spectra Physics argon-laser, a Spex Spectrometer with 850~mm
focal length and a liquid nitrogen cooled Ge-detector were employed.
In both set-ups, the samples were placed in a He-cooled flow
cryostat, which allows to cool the samples to 5~K. For adjusting the
temperature a silicon diode and a Lakeshore 331 temperature
controller were employed.

\subsubsection{Magnetic characterization}
The magnetic properties have been investigated in a home made 1~T
SQUID magnetometer working in the temperature range 5~-~330~K. The
samples for magnetic investigations are typically cut into
$9\times3$ or $5\times5$ mm$^2$ pieces, for in-plane and both in-
and out-of-plane comparative measurements, respectively. We study
both the temperature dependence of the magnetization at a constant
field and the sample response to the variation of the external field
at a constant temperature. In this case, we find it sufficient to
sweep the field in one direction only -- a half hysteresis loop --
and to obtain the second half by means of numerical reflection of
the data. Selected measurements of the full hysteresis loop have
been carried out in order to ensure the reliability of this
experimental procedure. Also, all the magnetization data presented
in the paper are already corrected for the diamagnetic signal from
the substrate, reflecting, in this way, the properties of the
investigated layers only. A detailed discussion of the substraction
procedure and related experimental difficulties specific to magnetic
measurements on thin films of diluted magnetic semiconductors and
the precautions undertaken in the present work are outlined in the
Appendix.
\subsubsection{Transmission electron microscopy}
Transmission electron microscopy studies have been carried out
on cross-sectional samples prepared by standard mechanical polishing
followed by Ar$^{+}$ ion milling at 4~kV for about 1~h. Conventional
diffraction contrast images in bright-field imaging mode and
high-resolution phase contrast pictures were obtained from a JEOL
2011 Fast TEM microscope operating at 200~kV and capable of an
ultimate point-to-point resolution of 0.19~nm and allowing to image
lattice fringes with a 0.14~nm resolution. The energy dispersive
x-ray analysis has been performed \emph{via} an Oxford Inca
EDS equipped with a silicon detector.
\section{Sample parameters}
In Table I  we summarize the data related to the GaN:Fe, (Ga,Fe)N
and reference GaN samples considered in this work. As already
mentioned, all samples have been grown on c-sapphire substrates. The
Fe-doped structures consist of a nominally 500~nm thick GaN:Fe or
(Ga,Fe)N layer deposited onto a 1~$\mu$m thick GaN buffer, whereas
the reference samples are mere 1~$\mu$m thick GaN buffers (nominally
undoped) on sapphire. In the table, the following values are listed:
the Cp$_{2}$Fe flow rate employed to grow the TM-doped layers, the
FWHM of the (0002) reflex from GaN determined by \emph{ex-situ}
HRXRD, the Fe$^{3+}$ concentration according to SQUID data and
evaluated from the Curie constant (in the case of low Fe
concentration, where the low-temperature paramagnetic contribution
is not covered by the ferromagnetic signal) and from the saturation
value of the ferromagnetic magnetization, respectively. The sample
numbers in the figures throughout the paper refer to Table I.

\begin{table*}
\caption{\label{tab:table1} Data related to
the investigated GaN, GaFe:N and (Ga,Fe)N samples.
    The following values are listed: the Cp$_2$Fe flow rate
    employed to grow the Fe-doped layers, the FWHM of the
    (0002) reflex from GaN determined by {\em {ex-situ}} HRXRD,
     the Fe concentration according to the Curie constant
     from SQUID data (in the case of low Fe concentration,
      where the Curie component is not covered by the
      ferromagnetic signal; S =5/2 and g = 2 is assumed) and from the saturation value of
      ferromagnetic magnetization $M_S$, assuming $S = g =1$.}
\begin{ruledtabular}
\begin{tabular}{cccccc|cccccc}
& & & & & & & & & & &\\
 sample & Cp$_{2}$Fe & FWHM & Fe$^{3+}$ conc. & $M$$_{S}$/$\mu$$_{B}$ & & sample & Cp$_{2}$Fe & FWHM & Fe$^{3+}$ conc.& $M$$_{S}$/$\mu$$_{B}$ &\\
 number & flow rate & [arcsec] & Curie comp.& [emu/(cm$^{3}$$\mu$$_{B}$)] & & number & flow rate & [arcsec] & Curie comp.& [emu/(cm$^{3}$$\mu$$_{B}$)] &\\
 & [sccm] &  & [cm$^{-3}$] &  & & & [sccm] & & [cm$^{-3}$] & &\\
& & & & & & & & & & &\\
\hline
& & & & & & & & & & &\\
\#310 & 0 & 265 & {---} & {---} & & \#467 & 175 & 279 & {---} & 3.02~$\times$~10$^{19}$ &\\
\#486 & 0 & 220 & {---} & {---} & & \#320 & 200 & 323 & 4.5~$\times$~10$^{19}$ & 1.05~$\times$~10$^{19}$ &\\
\#329 & 0 (GaN:Mg) & 287 & {---} & {---} & & \#470 & 200 & 258 & {---} & 2.26~$\times$~10$^{19}$ &\\
\#317 & 50 & 289 & 1.8~$\times$~10$^{19}$ & {---} & & \#472 & 225 & 282 & {---} & 2.37~$\times$~10$^{19}$ &\\
\#367 & 50 & 272 & 1.2~$\times$~10$^{19}$ & 8.41~$\times$~10$^{18}$ & & \#469 & 250 & 269 & {---} & 6.15~$\times$~10$^{19}$ &\\
\#371 & 50 & 220 & 1.4~$\times$~10$^{19}$ & 0 & & \#473 & 275 & 307 & {---} & 5.82~$\times$~10$^{19}$ &\\
\#409 & 50 & 249 & {---} & 1.2~$\times$~10$^{19}$ & & \#471 & 300 & 316 & {---} & 7.66~$\times$~10$^{19}$ &\\
\#318 & 100 & 288 & 1.9~$\times$~10$^{19}$ & {---} & & \#468 & 325 & 320 & {---} & 8.84~$\times$~10$^{19}$ &\\
\#319 & 150 & 308 & 2.3~$\times$~10$^{19}$ & 6.47~$\times$~10$^{18}$ & & \#466 & 350 & 302 & {---} & 8.95~$\times$~10$^{19}$ &\\
\#465 & 150 & 294 & {---} & {---} & & \#408 & 400 & 305 & {---} & 2.25~$\times$~10$^{20}$ &\\
& & & & & & & & & & &\\
\end{tabular}
\end{ruledtabular}
\end{table*}

\section{Experimental results and discussion}
\subsection{Transmission electron microscopy}
\label{sec:TEM}

\begin{figure}[htbp]
	\includegraphics[width=6.5cm]{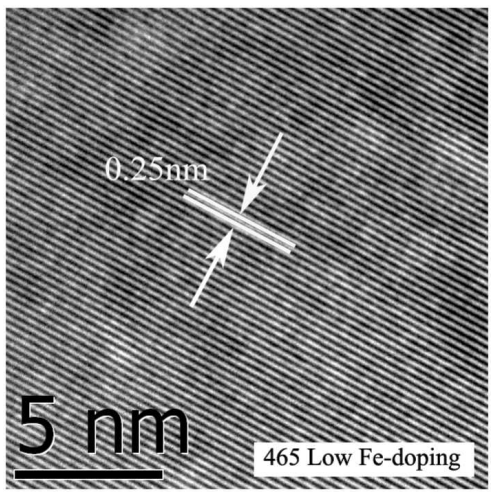}
		\caption{\label{fig:465} HRTEM lattice fringe image of the (0002) planes
demonstrating that low Fe-doping (sample \#465) does not cause detectable strain, the
lattice parameter ($d_{0002}=0.25\pm0.01$ nm) being the same as that of
undoped GaN.}
\end{figure}

\begin{figure}[htbp]
	\includegraphics[width=9.0cm]{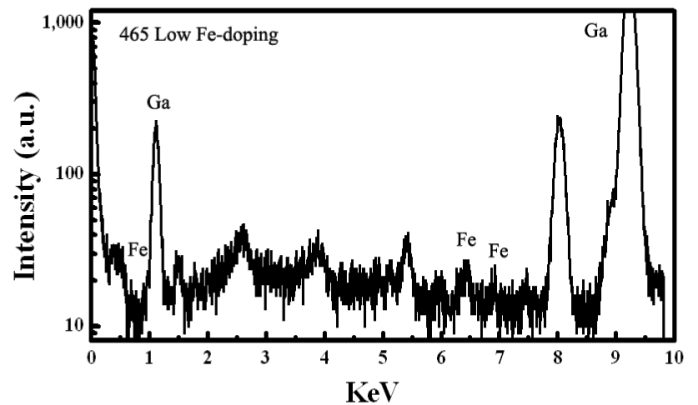}
		\caption{\label{fig:465EDX} EDS spectrum showing
the presence of Fe in the measured regions of the GaN layers doped with a
low concentration of Fe ions (sample \#465).}
\end{figure}

According to our TEM studies, in the limit of low Fe-doping, {\em
i.e.}, for Cp$_{2}$Fe flow rates in the range 50~--~150~sccm the
structure of the GaN matrix is not detectably affected. As indicated
in Fig.~\ref{fig:465}, the lattice spacing between the (0002) planes
is $0.250\pm0.010$~nm, perfectly matching that of undoped GaN. In
addition, the EDS spectra do not point to Fe aggregation
(Fig.~\ref{fig:465EDX}), indicating that the Fe ions are randomly
distributed over the lattice, presumably in substitutional sites of
Ga. Above 175 sccm, however, TEM studies reveal the presence of
nanocrystals embedded in the host GaN.

\begin{figure}[hbtp]
	\includegraphics[width=8.6cm]{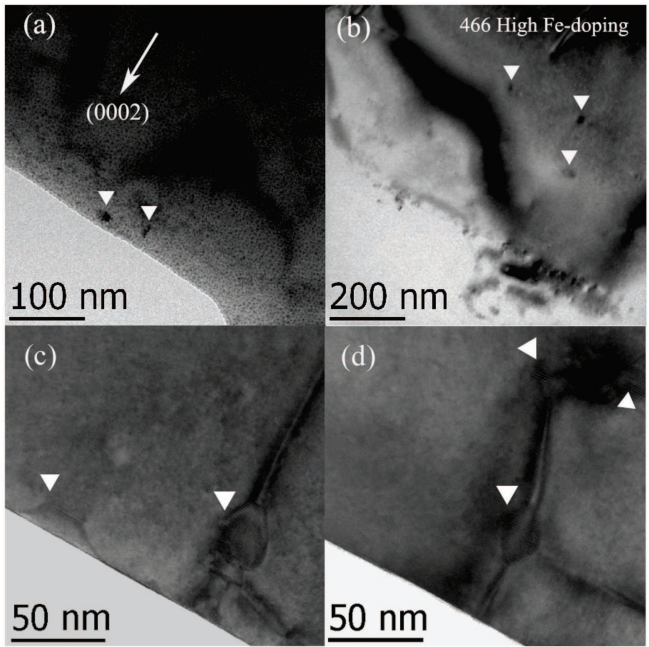}
		\caption{\label{fig:bf466} Bright-field transmission electron
micrographs of (Ga, Fe)N demonstrating the presence of precipitates
as indicated by the white triangles (a) close to the surface and
(b) in the mid of the epilayer. Some precipitates (c) are embedded
into the surface and others, (c) and (d) sink to the dislocations.}
\end{figure}

\begin{figure}[htbp]
	\includegraphics[width=8.6cm]{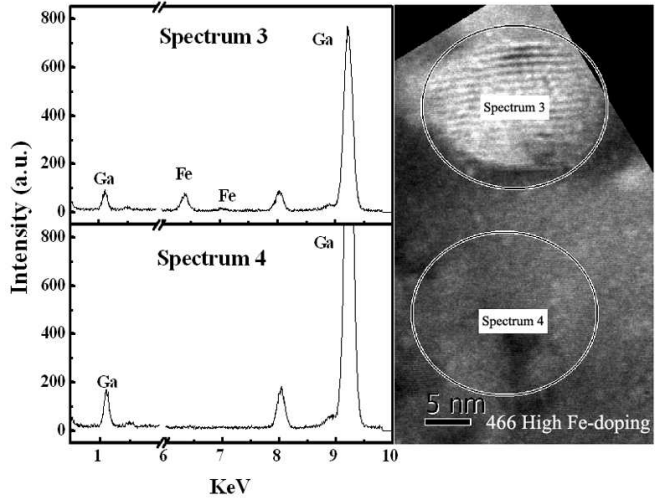}
		\caption{\label{fig:edxcompare} EDS spectra for the sample \#466 (left panel)
taken around the precipitate (right panel) showing that the concentration of
Fe is significantly enhanced in the region of the precipitate (recorded as
spectrum 3) as compared with the surrounding matrix (recorded as spectrum 4).}
\end{figure}

According to the TEM micrographs presented in
Figs.~\ref{fig:bf466}(a) and (b), the precipitates are located both
close to the surface and inside the epilayer in dislocation-free
regions, suggesting that precipitation is not related to the
presence of dislocations. On the other hand, Figs.~\ref{fig:bf466}(c)
and (d) prove that some nanocrystals are embedded into the upper
surface layers, while some others are decorating the screw- or
mixed- type threading dislocations in a volume proximal to the
surface.  Furthermore, EDS spectra demonstrate that the
concentration of Fe ions is largely enhanced within the
nanocrystals. As shown in Fig.~\ref{fig:edxcompare}, the Fe-$K_{a1}$
peak at 6.404 KeV can be seen clearly in the spectrum taken in the
region of a precipitate, while in the adjacent areas, the presence
of Fe is not detected. It should be mentioned that the precipitates can
be found only in the upper region of the layer within 300~nm from
the surface, and thus well above the interface between the nominally
undoped GaN buffer and the Fe-doped layer. The segregation of
TM-rich nanocrystals towards the surface of the samples, seems to
represent a general phenomenon in highly TM-doped MOCVD III-V compounds: a
similar effect has been observed for MnAs nanocrystals in MOCVD
grown (Ga,Mn)As.\cite{Lampalzer:2003_a}

\begin{figure}[htbp]
	\includegraphics[width=8.6cm]{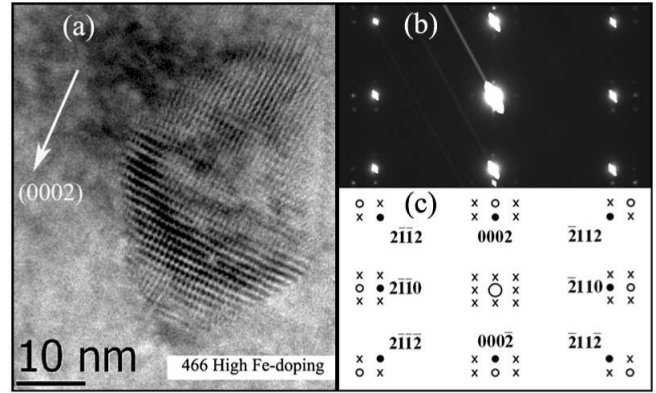}
		\caption{\label{fig:elongated}(a) Elongated precipitate observed in an HRTEM image
of Moir$\acute{e}$ fringes contrast. (b) SADP pattern acquired in the region around
the precipitate along the $(10\overline{1}0)$ zone axis; (c) the corresponding
schematic graph for indexing of the diffraction spots.}
\end{figure}

We estimate that the nanocrystals occupy less than 1\% of the film
volume and their diameter varies from $~$5 to$~$50~nm. Some
precipitates are elongated in the growth direction, as shown in
Fig.~\ref{fig:elongated}(a), such a behavior being expected within
the spinodal decomposition scenario\cite{Sato:2006_b}. This
non-spherical shape may lead to an additional magnetic anisotropy
that can elevate the blocking temperature. In
Fig.~\ref{fig:elongated}(b) the selected area diffraction pattern
(SADP) along the $(10\overline{1}0)$ zone axis taken around the
precipitate in Fig.~\ref{fig:elongated}(a) is shown. It reveals the
presence of an additional hexagonal phase with almost the same
orientation as the GaN matrix. In the corresponding schematic
indexed pattern  (Fig.~\ref{fig:elongated}(c)) closed circles
correspond to GaN, open circles to the additional phase and crosses
to the double diffraction between them. The lattice parameters of
this new phase are $a=0.268\pm 0.010$ nm, $c = 0.436 \pm0.010$ nm.
By comparing the lattice constants to those of Fe and iron nitride
nanostructures and thin films listed in Table~\ref{tab:table2}, we
tend to believe that $\varepsilon$-Fe$_{3}$N is formed in the
considered precipitates. However, it is also expected that the
GaN host can stabilize Fe$_x$N or (Ga,Fe)$_x$N nanocrystals in a
form not existing in the case of nanostructured Fe-N material
systems. Actually, such a situation has recently been revealed in
the case of MBE grown (Ge,Mn), where spinodal decomposition into Ge
and novel Mn-rich (Ge,Mn) nanocrystals has been
found.\cite{Jamet:2006_a}

\begin{figure}[htbp]
	\includegraphics[width=7.5cm]{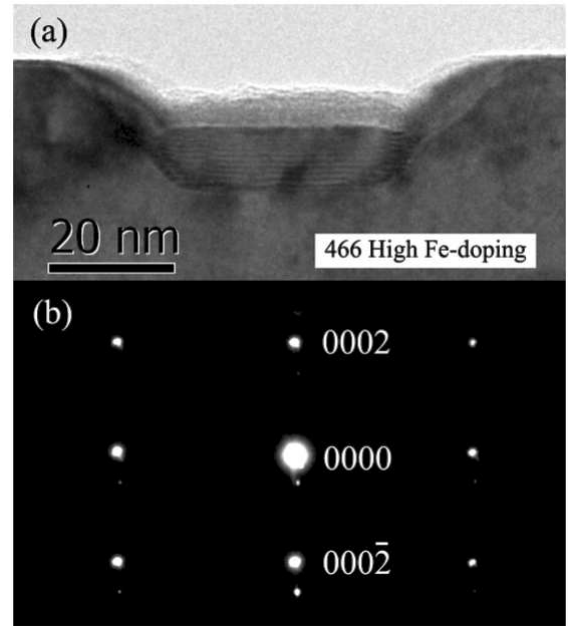}
		\caption{\label{fig:surface} (a) HRTEM image of a precipitate residing on the
surface of an highly Fe-doped sample (\#466); (b) the corresponding SADP pattern along
the $(10\overline{1}0)$ zone axis.}
\end{figure}

Our HRTEM studies allow to detect also precipitates residing at the
surface of the highly Fe-doped samples, as shown in
Fig.~\ref{fig:surface}(a). The SADP pattern in
Fig.~\ref{fig:surface}(b) suggests that the Moir$\acute{e}$ fringes in
this precipitate (Fig.~\ref{fig:surface}(a)) are formed by the interference
of the GaN (0002) planes of the matrix crystal with a set of lattice
planes parallel to (0002) with a \textit{d}-spacing of 0.204 $\pm$
0.010 nm, which is almost the same as that of the (111) planes of
$\gamma$-Fe or of the (110) planes of $\alpha$-Fe, as reported in
Table~\ref{tab:table2}. Therefore, these precipitates segregating to
the surface could consist of pure iron, this assumption being
consistent with the evaporation of nitrogen from the surface during
the MOCVD process and with the consequent hindering of the
Fe-nitride compounds formation.
\begin{table}
\caption{\label{tab:table2}Structural and magnetic parameters of
some iron and iron-nitride phases in nanostructures and thin films.
To be compared with our experimental values.}
\label{tab:table2}
\begin{ruledtabular}
\begin{tabular}{cccccc}
 &&\multicolumn{2}{c}{lattice parameter}\\&&\multicolumn{2}{c}{-----------------}\\&structure&$a$(nm)&$c$(nm)&
$\mu_\emph{\emph{B}}$ \\
\hline \\FeN& ZB & $0.430$\footnotemark[1] & -- &
AF or 0\footnotemark[1]  \\
& RS & 0.40\footnotemark[2] & -- &
--  \\\\
$\gamma'$-Fe$_{4}\emph{\emph{N}}$ & RS & $0.382$\footnotemark[2] &
-- &
2.21 \footnotemark[1] \\\\
$\gamma''$-$\emph{\emph{FeN}}_{0.91}$& ZB &
0.433\footnotemark[1] & -- &
1.7 or 0\footnotemark[1]  \\\\
$\varepsilon-\emph{\emph{Fe}}_{3}\emph{\emph{N}}$& WZ & 0.270
\footnotemark[3]& 0.433\footnotemark[3] &
1.9\footnotemark[4] \\\\
$\gamma'''$-$\emph{\emph{FeN}}_{0.5-0.7}$& RS &
0.450\footnotemark[1] & -- &
ferro\footnotemark[1]  \\\\
$\varsigma-\emph{\emph{Fe}}_{2}\emph{\emph{N}}$& HCP & 0.443\footnotemark[1] &
0.484\footnotemark[1] & para/1.5\footnotemark[8]    \\\\
$\gamma$-Fe& ZB & 0.361-0.370\footnotemark[5] & -- &
0.3-2.0\footnotemark[5]  \\\\
$\alpha$-Fe& BCC & 0.286\footnotemark[6] &
-- &
2.2\footnotemark[7]  \\\\
\end{tabular}
\end{ruledtabular}
\footnotetext[1]{Reference \onlinecite{Eck:1999_a} and references therein}
\footnotetext[2]{Reference
\onlinecite{Gallego:2005_a}}\footnotetext[3]{Reference \onlinecite{Gajbhiye:2004_a}}
\footnotetext[4]{Reference \onlinecite{Gallego:2004_a}}
\footnotetext[5]{Reference \onlinecite{Keavney:1995_a}}
\footnotetext[6]{Reference \onlinecite{Jiang:2004_a}}
\footnotetext[7]{Reference \onlinecite{Shi:1996_a}}
\footnotetext[8]{Reference \onlinecite{Rechenbach:1996_a}}
\end{table}

One may expect that the highly non-uniform Fe distribution over the
layer volume shown by TEM investigations will have a dramatic
influence on the electronic and magnetic properties of (Ga,Fe)N.
Experimental results presented in the next sub-sections confirm this
conjecture.

\subsection{Electrical properties}

It has been found previously that Fe doping reduces the electron
concentration in GaN and leads to a semi-insulating
material.\cite{Bougrioua:2005_a,Heikman:2002_a,Polyakov:2005_a} In
order to find out how the electronic properties evolve when the Fe
concentration surpasses the solubility limit, we have undertaken
electrical and optical measurements. Together with EPR spectroscopy,
such studies allow us to assess the charge and spin state of Fe ions
controlling the magnetic properties of the material system under
investigation. Hall effect and resistance measurements have been
carried out in the Van der Pauw geometry on $5\times5$~mm$^2$
square cuts of GaN, GaN:Fe, and (Ga,Fe)N wafers, whose
characteristics are summarized in Table I. Soldered indium contacts
are employed and the ohmic characteristics of all contacts are
carefully tested. The Hall resistance has been measured in the
magnetic field $B$~=~$\pm 500$~mT using currents in the range 10 to
200~$\mu$A. The employed current values are high enough to achieve
high Hall voltages for better accuracy, but still sufficiently low
to avoid Joule heating of the sample.

\begin{figure}[htbp]
    \centering
        \includegraphics[width=8.6cm]{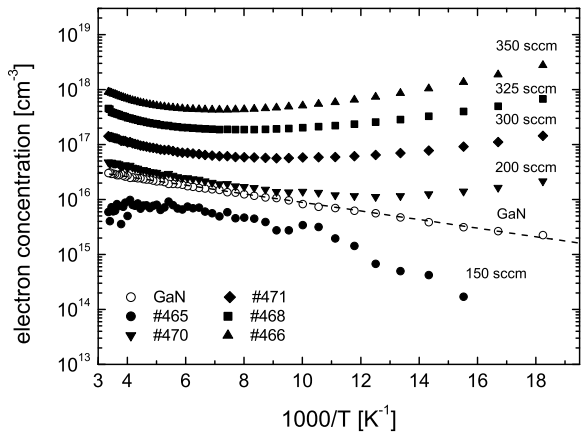}
    	\caption{Electron concentration $n$ from Hall effect measurements
    as a function of the inverse temperature for a GaN layer and a series of
    GaN:Fe -- (Ga,Fe)N samples with differing Fe content controlled by the Cp$_2$Fe
    flow rate.}
    \label{Figure:n_vs_T_corr_1}
\end{figure}

\begin{figure}[htbp]
    \centering
        \includegraphics[width=9.0cm]{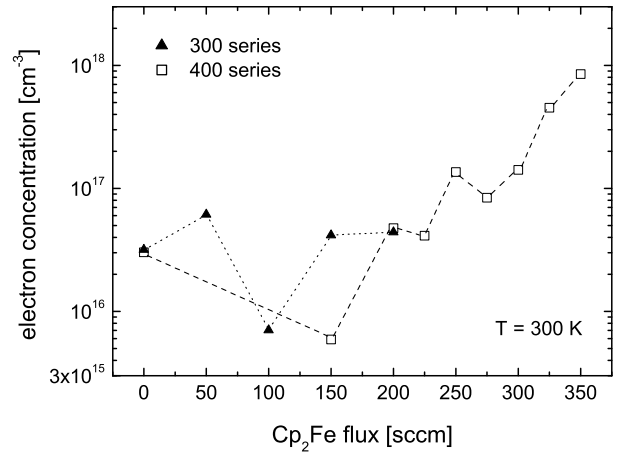}
    \caption{Electron concentration $n$ from Hall effect measurements as a
    function of the Cp$_2$Fe flow rate for two GaN reference layers and for
    two series of (GaN:Fe -- Ga,Fe)N samples (triangles and squares, respectively) at
    room temperature. Reduced electron concentrations in the
    low flow rate region points to a semi-insulating character of Fe-doped layers
    in this regime.}
    \label{Figure:n_vs_flux_corr_1}
\end{figure}

In the studied samples, both the (Ga,Fe)N film and the GaN buffer
layer are conducting, so that a procedure suitable for the
evaluation of electrical characteristics in the presence of a
parallel conducting channel has to be employed. Accordingly, in
order to obtain the conductance tensor components $G_{ij}^{(f)}$ of
the (Ga,Fe)N films, the buffer contribution $G_{ij}^{(b)}$ has to be
subtracted from the total conductance $G_{ij}$. We evaluate
$G_{ij}^{(b)}$ and $G_{ij}^{(f)}$ from the measured values of the
square resistance $R$ and Hall resistance $R_{\mathrm{Hall}}$ for
the reference GaN sample and for the samples containing the (Ga,Fe)N
films according to:
\begin{equation}
G_{xx}^{(b)} = t_bR^{(r)}/[t_r({R^{(r)}}^2 + {R_{\mathrm{Hall}}^{(r)}}^2)];
\end{equation}
\begin{equation}
G_{yx}^{(b)} = t_bR_{\mathrm{Hall}}^{(r)}/[t_r({R^{(r)}}^2 + {R_{\mathrm{Hall}}^{(r)}}^2)];
\end{equation}
\begin{equation}
G_{xx} = R/[(R^2 + R_{\mathrm{Hall}}^2)];
\end{equation}
\begin{equation}
G_{yx} = R_{\mathrm{Hall}}/[(R^2 + R_{\mathrm{Hall}}^2)];
\end{equation}
where $t$, $t_b$, and $t_r$ are the thickness of the (Ga,Fe)N film,
of the buffer layer, and of the reference sample, respectively.
Therefore, the (Ga,Fe)N conductance is $G_{ij}^{(f)} = G_{ij} -
G_{ij}^{(b)}$ and from that we obtain the resistivity $\rho$ and the
Hall electron concentration $n$ of the (Ga,Fe)N films in the
standard way:
\begin{equation}
\rho = t/G_{xx}^{(f)}(B = 0);
\end{equation}
\begin{equation}
n = B{(G_{xx}^{(f)}}^2 + {G_{xy}^{(f)}}^2)/(etG_{xy}^{(f)}).
\end{equation}
We must point out that within this model we assume the carrier
concentration of the GaN buffer layer to be the same in all the
measured samples and that there is no diffusion of Fe into the
buffer. Fluctuations in the buffer properties will affect the data
in a significant way when the conductance of the (Ga,Fe)N layer
becomes small. All the studied samples are found to be n-type with
R$_{\mathrm{H}}< 0$, where
R$_{\mathrm{H}}$~=~R$_{\mathrm{Hall}}$$\cdot$ t/B is the Hall
constant. As shown in Fig.~\ref{Figure:n_vs_T_corr_1}, in the region
of low Cp$_2$Fe flow rate we find the magnitude and temperature
dependencies of the Hall concentration typical for MOCVD GaN
epitaxial films, in which electrons freeze-out at residual donors at
low temperatures. The obtained activation energy of the carriers
in GaN, namely $E_a = 31$~meV, supports this conclusion.
According to our findings summarized in
Figs.~\ref{Figure:n_vs_T_corr_1} and \ref{Figure:n_vs_flux_corr_1},
the samples grown at the low Cp$_2$Fe flow rate of 150~sccm show
reduced Hall concentration as compared to GaN, which may result from
trapping of the electrons by Fe ions, $\mbox{Fe}^{3+} \rightarrow
\mbox{Fe}^{2+}$, an effect consistent with the position of the
Fe$^{3+}$/Fe$^{2+}$ acceptor level in the band gap of GaN, as we
will discuss in more detail in the next subsection.

\begin{figure}[htbp]
    \centering
        \includegraphics[width=8.6cm]{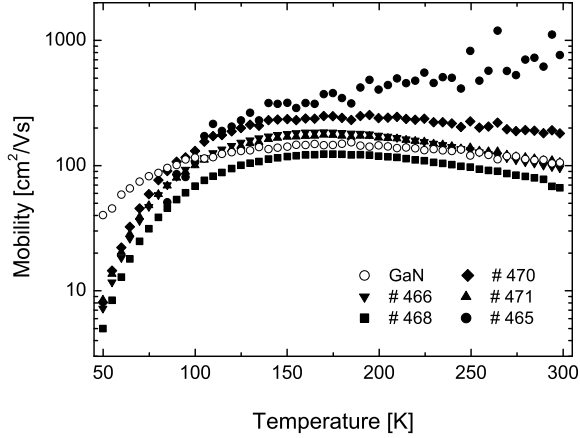}
    \caption{Electron Hall mobility $\mu$ as a function of temperature for a GaN
    layer and a series of (Ga,Fe)N samples with differing Fe content controlled
    by the Cp$_2$Fe flow rate.}
    \label{Figure:mu_vs_T}
\end{figure}

Interestingly, when the Cp$_2$Fe flow rate exceeds 150~sccm, which
according to SIMS data corresponds to a Fe content above
$10^{20}$~cm$^{-3}$, we observe a strong increase of the Hall
concentration, and a change in the character of its temperature
dependence, a behavior clearly seen in
Figs.~\ref{Figure:n_vs_T_corr_1} and \ref{Figure:n_vs_flux_corr_1}.
For the highest flow rate of 350~sccm, the Hall concentration
reaches $10^{18}$~cm$^{-3}$, a value more than two orders of
magnitude higher than that for the sample grown with the Cp$_2$Fe
flow rate of 150~sccm. This effect might suggest the appearance of
an additional conductance channel associated with the hopping of
electrons between Fe ions at large magnetic ions densities.
Actually, hopping conductance involving Fe impurities was discussed
for InP,\cite{Gorodynskyy:2004_a} where, however, the measured
resistance was found to be much larger than in our case. Indeed, the
relatively large values for the Hall mobility $\mu = 1/(en\rho)$
shown in Fig.~\ref{Figure:mu_vs_T} make the hopping scenario
improbable.  We also note that, as discussed in
subsection~\ref{sec:TEM}, Fe-rich nanocrystals are formed at higher
flow rates, so that we deal with carrier transport in a highly
inhomogenous medium. Under such conditions, the relation between the
Hall and carrier densities becomes intricate. It is more probable
that (at least a part) of the rise in the Hall concentration is
caused by an increase in the oxygen donor density for large Cp$_2$Fe
flow rates. SIMS analysis shows that the oxygen concentration
increases by over two orders of magnitude upon changing the flow
rate from 50 to 350~sccm and at the highest flow rate it becomes
comparable to the iron concentration. Finally, it is also possible
that some part of the Fe ions occupies interstitial positions in
which Fe$^{2+}$ acts as a double donor. Optical and ESR data
presented in the next subsections show that the charge state of Fe
ions transforms from +3 to +2 with increasing Fe concentration. This
effect supports the scenario which assigns the enhancement in the
carrier concentration to the increase of the donor concentration with
the Fe content, though the origin of the relevant impurity or defect
remains uncertain. We will return to this open question when
discussing the results of the EPR measurements.

\subsection{Optical properties}
\begin{figure}[htbp]
    \centering
        \includegraphics[width=8.6cm]{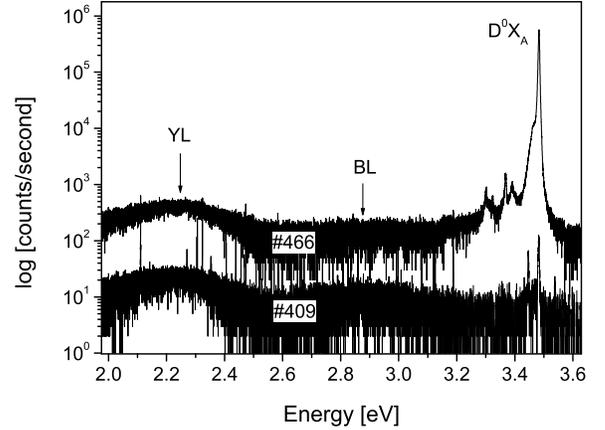}
    \caption{Photoluminescence spectra at 10~K: comparison between
    samples with low (\#409) and high (\#466) Fe concentration.}
    \label{fig:PL_all_1}
\end{figure}

\begin{figure}[htbp]
    \centering
        \includegraphics[width=8.6cm]{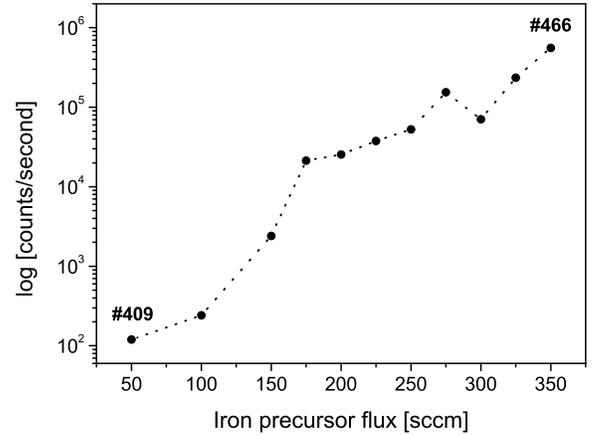}
    \caption{Near-band-edge emission as a function of the Fe precursor flow rate.}
    \label{fig:NBBE_1}
\end{figure}

Photoluminescence measurements in the infrared, visible, and
ultra-violet regime have been carried out on the whole series of
samples with increasing Fe concentration. Figure~\ref{fig:PL_all_1}
presents PL spectra acquired in the energy range 2.0~--~3.6~eV at
10~K for two GaN:Fe samples grown with iron precursor flow rates of
50 and 350~sccm. In both samples a weak defect-related yellow
luminescence band (YL) centered around 2.2~eV is seen, as well as the
excitonic near-band-edge emission. The intensity of the latter is
found to increase by four orders of magnitude between the samples
with the lowest and the highest Fe content, as shown in
Figure~\ref{fig:NBBE_1}. The peak position varies between 3.4811 and
3.4833~eV at 10 K, depending on the layer thickness. The determined
position and thermal behavior is identical to that of the
neutral-donor-bound exciton D$^{_0}$X$_A$, related to the upper
valence subband of wz-GaN and observed in nominally undoped
GaN.\cite{Wegscheider:2006_a} However, the peak is noticeably wider,
with a FWHM of 4.65~meV for the lowest iron concentration and broadens
with increasing doping level.

\begin{figure}[htbp]
    \centering
        \includegraphics[width=8.6cm]{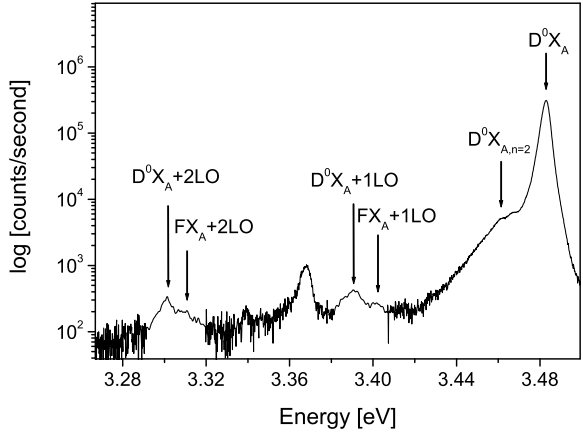}
    \caption{Near band edge photoluminescence in GaN:Fe. The positions of D$^{0}$X$_A$
    and the two electron satellite D$^{0}$X$_{A,n=2}$ are indicated by arrows, as well
    as those of the LO phonon replicas for the bound- and free-exciton (FX$_A$).}
    \label{fig:PL_twoe_1}
\end{figure}

The lower intensity emission at 3.4609~eV visible in
Fig.~\ref{fig:PL_twoe_1} is assigned to a two electron satellite
D$^{0}$X$_{A,n=2}$, where the recombination of the D$^{0}$X$_A$
leaves the donor in its excited $n = 2$
state.\cite{Korona:2002_a,Wysmolek:2002_a} From the separation
between the principal D$^{0}$X$_A$ and the D$^{0}$X$_{A,n=2}$
maximum, the donor-ionization energy can be
obtained.\cite{Monemar:2001_a} In our case it results in
E$_D$~$\approx$~30 meV, pointing to oxygen or silicon as principal
binding site for the D$^{0}$X$_A$. An increase in the oxygen
concentration with increasing iron content is confirmed by SIMS
data. However, due to the broadening of the excitonic transitions
the presence of another donor, with similar binding and ionization
energies cannot be ruled out.

Quenching of the excitonic luminescence upon doping with iron is
expected, since transition metal impurities in III-V semiconductors
are well known to act as carrier lifetime
killers.\cite{Pantelides:1992_a} The recovery of the luminescence
intensity with increasing iron content appears, therefore,
remarkable and cannot be explained by the increase in shallow donor
concentration alone. Apparently, the neutral donors compete
successfully with Fe$^{2+}$ in the hole capture. The strong quenching of
the excitonic emission observed at low iron contents seems to
be predominantly due to the high electron capture rate  of
Fe$^{3+}$.  The change of the charge state of iron with the rise of
the electron concentration in our system is, hence, responsible for
the recovery of the near band edge emission.

\begin{figure}[htbp]
    \centering
        \includegraphics[width=8.6cm]{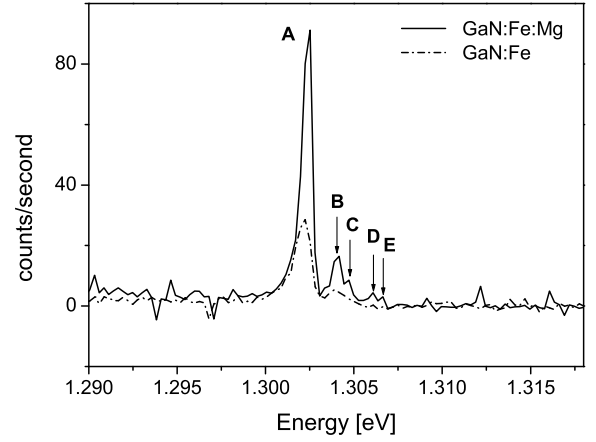}
    \caption{Intensity dependence of the Fe$^{3+}$ internal transition
    as a function of co-doping with Mg.}
    \label{fig:PL_FeMg_1}
\end{figure}

\begin{figure}[htbp]
    \centering
        \includegraphics[width=8.6cm]{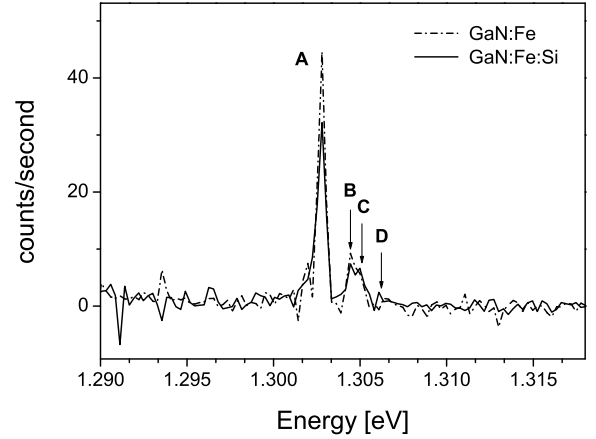}
    \caption{Intensity dependence of the Fe$^{3+}$ internal
    transition as a function of co-doping with Si.}
    \label{fig:PL_FeSi_1}
\end{figure}

The recombination of electrons and holes on the Fe$^{2+/3+}$
acceptor level is accompanied by the characteristic intra-center
transition of Fe$^{3+}$. Figures~\ref{fig:PL_FeMg_1} and
\ref{fig:PL_FeSi_1} depict the zero phonon lines of the internal,
spin forbidden $^{4}$T$_1$(G)-$^{6}$A$_1$(S) transition   near
1.3~eV, measured at 20~K under above band-gap excitation in three
samples with the same iron content (grown at the lowest iron
precursor flow rate) but different electron concentration. The peaks
labelled B, C, D and E at the high energy tail of peak A stem from
higher lying levels of the $^{4}$T$_1$(G)
multiplet.\cite{Baur:1994_b,Heitz:1992_a,Malguth:2006_a}

It was found that co-doping with Si (Fig.~\ref{fig:PL_FeSi_1})
reduces the intensity of the Fe$^{3+}$ related emission, while
co-doping with Mg (Fig.~\ref{fig:PL_FeMg_1}) leads to an intensity
enhancement. This behavior is consistent with that of the near band
edge luminescence. The increased  electron concentration in
GaN:Fe,Si as compared to GaN:Fe leads to a  reduction of the carrier
recombination rate at the Fe$^{2+/3+}$ acceptor level and, thus, the
emission intensity, whereas reduction of the electron concentration
(by codopig with Mg) leads to the opposite effect.

\subsection{Electron paramagnetic resonance}
\begin{figure}[htbp]
    \centering
        \includegraphics[width=9.0cm]{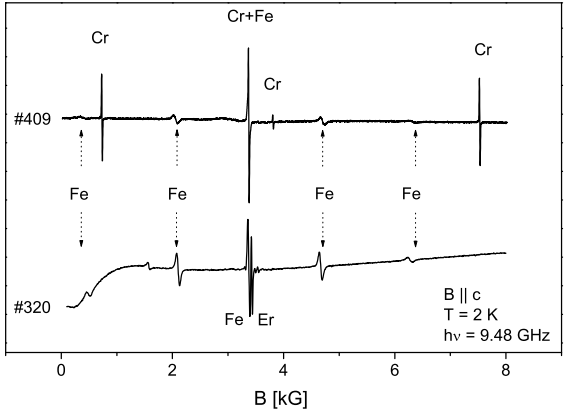}
    \caption{EPR spectra - first derivative of the microwave absorption with respect
    to the magnetic field - of Fe$^{3+}$ in GaN for the magnetic field direction close
    to the [0001] GaN axis. The contribution of Cr$^{3+}$ and Er$^{3+}$ contamination
    of the sapphire substrate is easily seen. The lower spectrum was taken at a higher
    microwave power than the upper one (since the EPR signal of Er$^{3+}$ is not as easily
    saturated as that of Cr$^{3+}$) and a background due to microwave absorption in the
    cryostat is observed.}
    \label{fig:EPR_Fe_Cr}
\end{figure}

The samples have also been characterized by electron paramagnetic
resonance. The advantage of this technique is that the contributions
to magnetization originating from different paramagnetic impurities
can be determined separately. It has been found that in all studied
structures, independent of the iron concentration, an EPR signal from
isolated, substitutional Fe$^{3+}$ ions is observed. In most samples
this signal is superimposed to a strong spectrum stemming from some
Cr$^{3+}$ contamination of the sapphire substrate. In other samples
(e.g., \#320) Er$^{3+}$ was observed in the substrate instead of
Cr$^{3+}$. Typical EPR spectra for the magnetic field oriented along
the c-axis are shown in Fig.~\ref{fig:EPR_Fe_Cr}.

\begin{figure}[htbp]
    \centering
        \includegraphics[width=9.0cm]{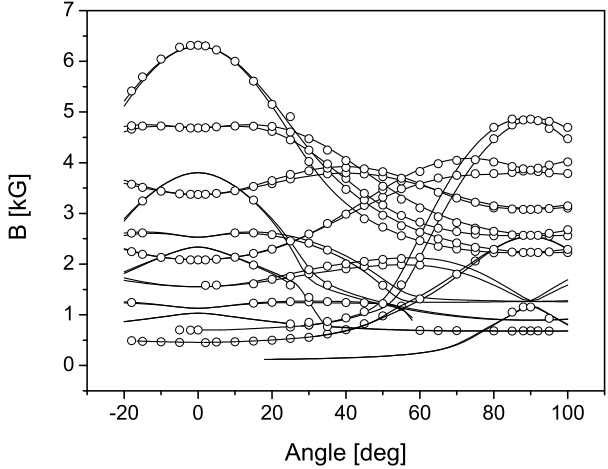}
    \caption{Angular dependence of the EPR line positions for sample \#406.
    The solid lines represent the fit to experimental data with the Hamiltonian
    and parameters given in the text.}
    \label{fig:EPR_ang_dep}
\end{figure}

The spin Hamiltonian parameters for the Fe$^{3+}$ ion are determined
by a least squares fit of the angular dependence of the EPR line
positions for the magnetic field {\bf B} rotated in the (11$\bar2$0)
plane - as depicted in Fig.~\ref{fig:EPR_ang_dep} - with the spin
Hamiltonian for an $S=5/2$ ion in C$_{3v}$ symmetry:

\begin{equation}
H = \mu_{\mathrm{B}}{\mbox{\bf B}}g{\mbox{\bf S}}-
\frac{2}{3}B_{4}(O^{0}_{4}+20\sqrt{2}{O}^{3}_{4})+ B^{0}_{2}O^{0}_{2}+B^{0}_{4}{O}^{0}_{4};
\end{equation}

Here the first and second term describe the Zeeman- and the
cubic-crystal-field-interaction respectively, and the last two terms
represent the second and fourth order trigonal crystal field
interactions. The spin Hamiltonian parameters, especially the value
of the second order trigonal parameter $\left|B^{0}_{4}\right|$, are
found to vary slightly with the sample thickness, giving indication
that the layers are strained.

Typical parameters for a 1 $\mu$m thick layer at moderate Fe
precursor flow rates (50~--~200 sccm) are: $g_{||} = 2.009 \pm
0.002$, $g_{\bot} = 2.005 \pm 0.002$, $\left|B^{0}_{2}\right| =
237.2 \pm 0.6$~G, $B_4=0.11\pm 0.01$~G, and $\left|B^{0}_{4}\right|
= 0.8 \pm 0.1$~G.

\begin{figure}[htbp]
    \centering
        \includegraphics[width=9.0cm]{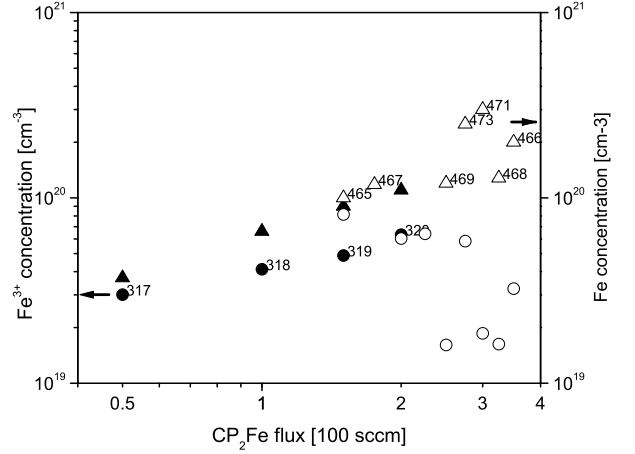}
    \caption{Concentration of Fe$^{3+}$ estimated from the EPR signal intensity
    (circles) as compared to the total Fe concentration measured by SIMS (triangles) vs.~the
    iron precursor flow rate. Full and open symbols identify two different sample series,
    grown at the same conditions.}
    \label{fig:EPR_vs_SIMS}
\end{figure}

Figure~\ref{fig:EPR_vs_SIMS} shows the concentration of Fe$^{3+}$
estimated from the EPR signal intensity (circles) compared to the
total Fe concentration measured by SIMS (triangles) as a function of
the iron source flow rate for two series of samples (full and open
symbols). The number of Fe$^{3+}$ ions in the samples has been
determined by comparison with a reference, phosphorus-doped Si
sample with a known P concentration of 2~$\times$~10$^{14}$
cm$^{-3}$. The measurements were performed at low microwave powers
in order to ensure that the intensities of both Fe$^{3+}$ and P
signals were far from the saturation regime. Since the samples could
not be measured simultaneously, the relative number of ions is
determined with an accuracy of 20\%, the uncertainty being due to
possible changes in the quality factor of the microwave cavity.
Moreover, the relative concentration contains an additional error
related to the assumed thickness of the doped layer: the nominal
thickness, obtained from the number of observed thickness
oscillations in ellipsometry can differ by 20\% from the one
estimated by SIMS. In the high Fe-doping regime, the accuracy is even
lower, since the surface roughness inhibits the observation of the
oscillations in ellipsometry as well as a reliable estimation of the
thickness from SIMS data. There is also a systematic error related
to the estimation of the P content, so that the absolute
concentration of Fe$^{3+}$ determined from the EPR intensity may
differ from the true value by not more than a factor of two. Despite
this uncertainty, a clear trend is visible in
Fig.~\ref{fig:EPR_vs_SIMS}: with increasing iron precursor flow rate -- up to
about 200~sccm -- the concentration of substitutional Fe$^{3+}$ ions
increases continuously, but at a lower rate than the total iron
concentration in the sample. Above 200~sccm the Fe$^{3+}$ content is
reduced and considerable fluctuations are observed from sample to
sample. We relate this effect to the change of the Fermi level
position and, hence, greater occupancy of the Fe$^{2+}$ charge
state. In the high doping regime a drastic increase of the oxygen
donor concentration, by about two orders of magnitude as compared to
that of GaN:Fe grown at low Cp$_{2}$Fe flow rates, is detected by
SIMS. The oxygen accumulation seems to be correlated with the
enhanced sample roughness: as previously underlined, at low flow rates (up to 200~sccm) the
GaN:Fe growth is two-dimensional (and the oxygen concentration stays
below 10$^{18}$ cm$^{-3}$), whereas above 200~sccm the (Ga,Fe)N
nucleation is granular with over 1 $\mu$m fluctuations in the grain
height at 350 sccm.

\begin{figure}[htbp]
    \centering
        \includegraphics[width=9.0cm]{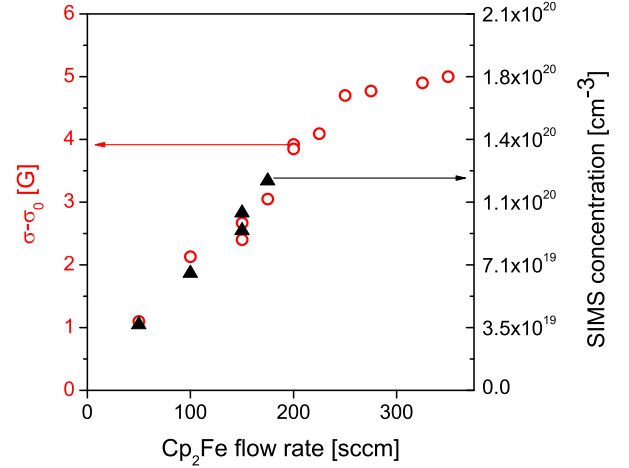}
    \caption{EPR linebroadening (left y-axis) as a function of the iron
    precursor flow rate (hollow circles) for $\sigma_{0}=12$~G. The SIMS
    concentration (right y-axis) at low flow rates, denoted by full triangles,
    is shown for comparison.}
    \label{fig:EPR_broad}
\end{figure}

The presence of Fe$^{2+}$ ions cannot be determined directly by EPR
since the ground state is a spin singlet state and there is no
resonance transition observed in the X-band. We have detected a
metastable 20\% increase of the EPR signal intensity of Fe$^{3+}$
under illumination with light of energies above 1.2 eV, confirming
the presence of Fe in the 2+ charge state, but no estimations of the
actual concentration can be made from such experiments. This
increase depends predominantly on the concentration of other, not
iron-related trap centers in the sample as well as on the carrier
capture rates.  Therefore, in order to estimate the total
concentration of substitutional Fe ions in the sample, we have
analyzed the EPR line width of Fe$^{3+}$. The line shape $A(B)$ of
Fe$^{3+}$ is found to be excellently described by the derivative of
the Gaussian error function:
\begin{equation}
A(B) = \frac{1}{\sqrt{2\pi}\sigma}\exp\left(-\frac{(B-B_{0})^{2}}{2\sigma^{2}}\right),
\end{equation}
where $\sigma^{2} =
\left\langle\left(B-B_{0}\right)^{2}\right\rangle$ is the second
moment of the distribution.  At very low concentrations, the
linewidth $\sigma$ should be mainly governed by interactions with
nuclear magnetic moments of the Ga and N isotopes, however, with
increasing Fe concentration dipole-dipole interactions between iron
ions should lead to line broadening. The broadening due to
dipole-dipole interaction consists of two contributions: one due to
the interaction with other 3+ ions and one related to the
interaction with Fe ions in the 2+ charge state. This is true as
long as the Fermi level is pinned to the  Fe$^{2+/3+}$ acceptor
level and the oxygen ions have no magnetic moment, which seems to be
justified, since no EPR transition related to O donors is observed.
Both contributions depend on the i$^{th}$ to j$^{th}$ ion distance
r$_{ij}$, as r$_{ij}^{-3}$. Assuming that the ions as well as their
charges are uniformly distributed in the lattice, one can expect a
linear correspondence between the dipolar broadening
$\sigma$-$\sigma_{0}$ and the Fe concentration.
Figure~\ref{fig:EPR_broad} gives the linewidth as a function of iron precursor
flow rate for the -1/2 $\rightarrow$ 1/2 transition of Fe$^{3+}$ at
{\bf B}~$\parallel \left[0001\right]$ for the same samples shown
already in Fig.~\ref{fig:EPR_vs_SIMS}.

As it can be seen, the linewidth increases continuously with the
Cp$_{2}$Fe flow rate up to 250~sccm and saturates above this value.
By taking $\sigma_{0} = 12$~G for the unbroadened linewidth we
obtain, in the low doping regime, exactly the same slope for the
dipolar line broadening versus Cp$_{2}$Fe flow rate as that for the
SIMS concentration in Fig.~\ref{fig:EPR_vs_SIMS}. This allows us to
set the solubility limit for substitutional iron in GaN at our
growth conditions (that is, the saturation concentration obtained
from the EPR linewidth) to $1.8\times 10^{20}$~cm$^{-3}$,
\textit{i.e.}, 0.4\%. It should be noted that in contrast to
concentration estimations from the EPR signal intensity, in the case
of EPR linewidth there is no error related to the determination of
the sample thickness, therefore the comparison with SIMS data is
more straightforward.

The relatively narrow linewidth in the high doping regime visible in
Fig.~\ref{fig:EPR_broad} indicates that the substitutional Fe ions
remain well isolated and interact only very weakly $\textit{via}$ dipole-dipole
interaction. This is consistent with the TEM data, revealing the
presence of iron-rich nanocrystals in which excess iron is
accumulated.  These nanocrystals do not contribute to the Fe$^{3+}$
EPR. As we already underlined, the nanocrystals are located in the
upper half of the (Ga,Fe)N layer towards the surface and may deplete
this part of the layer from isolated Fe ions which thus remain in
the lower region of the layer only.  It is worth also noting that
the coexistence of Fe$^{3+}$ and Fe$^{2+}$ ions implied by the EPR
data as well as by the SQUID results, suggests that the Fermi level
is pinned well below the conduction band minimum by the
Fe$^{3+}$/Fe$^{2+}$ acceptor states. Such pinning is in
contradiction with the relatively high value of electron
concentration and mobility we observe in this high Cp$_{2}$Fe flux
range, as shown in Figs.~\ref{Figure:n_vs_T_corr_1} and
\ref{Figure:mu_vs_T}, giving clear indication that the Fermi level
is located either in the conduction band or in the donor impurity
band.  The simultaneous observation of Fe$^{3+}$ ions (as evidenced
by EPR) and free electrons (as shown by the Hall data) in the high
Fe concentration regime is one of the surprising findings of the
present work. In view of our TEM data, it is natural to suggest that
the conduction proceeds $\textit{via}$ the upper region of the layer
depleted from isolated Fe acceptors, while Fe$^{3+}$ ions exist
mainly in the part of the layer closer to the interface with the GaN
buffer.

\subsection{Magnetic properties of (G$\normalfont\bf{a}$,F$\normalfont\bf{e}$)N}
\subsubsection{Paramagnetic contributions}

\begin{figure}[htbp]
    \centering
        \includegraphics[width=9.0cm]{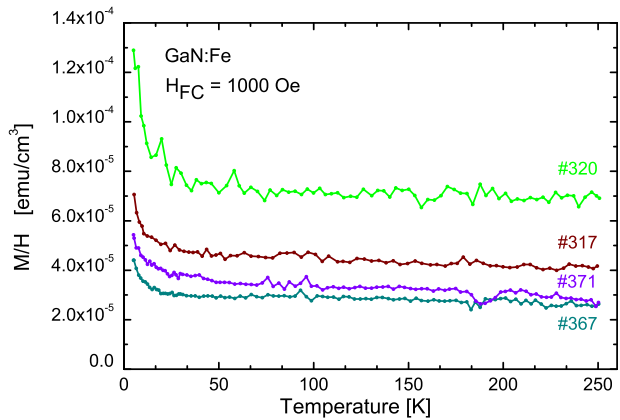}
    \caption{Magnetic susceptibility as a function of  temperature, showing the
    temperature independent and the temperature dependent components assigned to the
    Van Vleck and Curie paramagnetism, respectively. The ferromagnetic and substrate
    contributions (temperature independent and linear in the field) have been subtracted.}
    \label{fig:SQUID_low_para}
\end{figure}

\begin{figure}[htbp]
    \centering
        \includegraphics[width=8.9cm]{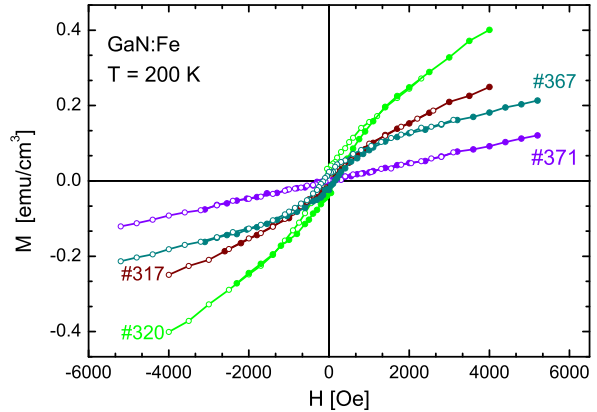}
\caption{Magnetization from SQUID  measurements at 200~K for a
series of (Ga,Fe)N samples with different Fe content (solid
symbols). The substrate contribution (temperature independent and
linear in the field) has been subtracted in this and in the figures
that follow. Data obtained by numerical reflection, if only a half
of the hysteresis cycle has been measured, are shown by open
symbols.}
    \label{fig:SQUID_low_ferrom}
\end{figure}

Figure~\ref{fig:SQUID_low_para} gives the temperature dependence of
the magnetic susceptibility $\chi(T)$ obtained from SQUID
magnetization measurements at a constant magnetic field of 0.1~T for
four GaN:Fe samples with a low Fe content. As it can be seen, two
components of $\chi(T)$ may be distinguished: one is a Curie-type of
paramagnetism $\chi(T) \sim 1/T$, which we assign to
Fe$^{3+}$(d$^{5}$) ions. The concentration of Fe$^{3+}$ ions
obtained under this assumption is shown in Table I for the samples
under investigation. The second component is a temperature
independent contribution dominating at high temperatures. However,
as shown in Fig.~\ref{fig:SQUID_low_ferrom}, the magnetization at
high temperature in addition to the paramagnetic part linear in the
magnetic field, contains a ferromagnetic component. Properties and
origin of the high temperature ferromagnetism in (Ga,Fe)N will be
discussed later.

We attribute the term linear in the magnetic field to Van Vleck
paramagnetism, most likely originating from the Fe ions in the 2+(d$^{6}$) charge state. The presence of Fe ions in such a charge
state seems to correlate with the Hall effect measurements which,
according to Fig.~\ref{Figure:n_vs_flux_corr_1}, indicate that the
Fermi level for low Fe flow rates may be pinned to the Fe$^{3+/2+}$
level. The coexistence of Fe in the 3+ and 2+ charge states is also
confirmed by photo-EPR results and the theoretical explanation of
the Van Vleck contribution is currently being
formulated.\cite{Kiecana:2006_a}

\subsubsection{Ferromagnetic properties}

\begin{figure}[htbp]
    \includegraphics[width=9.0cm]{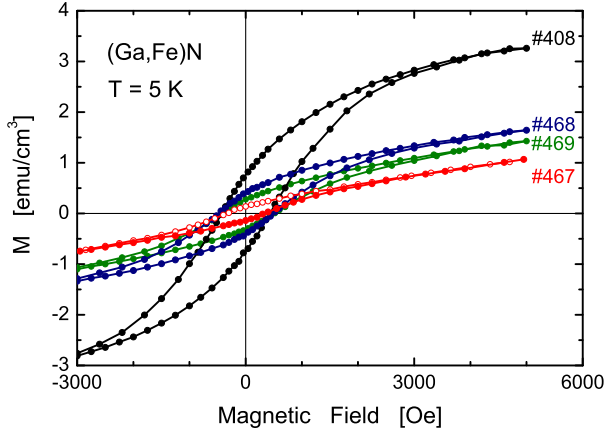}
        \caption{Magnetization  at 5~K for a series of (Ga,Fe)N samples with
various Fe contents (solid symbols). Data obtained by numerical
reflection, if only a half of the hysteresis cycle has been
measured, are shown by open symbols.} \label{Figure:SQUID_high_x_5}
\end{figure}

Ferromagnetic signatures, such as magnetic hysteresis, remanent
magnetization ($M_{\mathrm{rem}}$), and coercivity
($H_{\mathrm{c}}$) are detectable in all our Fe-doped films but with
the increase of the total Fe concentration above the solubility
limit of 0.4\% the ferromagnetic component exceeds the
paramagnetic contributions previously discussed. This is depicted in
Fig.~\ref{Figure:SQUID_high_x_5}, which shows the magnetization
cycles at 5~K for samples with various Fe contents. A rather clear
ferromagnetic component is seen, with its strength increasing upon
the increase of the Fe content. For the layers with the highest Fe
concentration, we
recover about 25\% of the Fe ions contributing to the saturation
value of magnetization, if $S = 2$ and $g = 2$ are assumed.

\begin{figure}[htbp]
    \includegraphics[width=8.6cm]{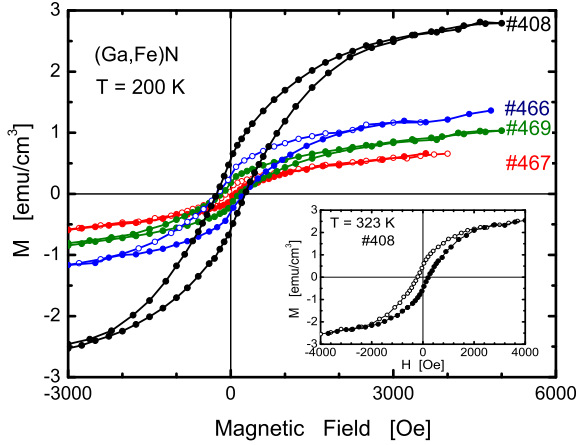}
        \caption{Magnetization at 200~K for a series of (Ga,Fe)N samples
with various Fe content (solid symbols). Inset: above room
temperature $M(H)$ for sample \#408. Data obtained by numerical
reflection, if only a half of the hysteresis cycle has been
measured, are shown by open symbols.}
    \label{Figure:SQUID_high_x_200}
\end{figure}

\begin{figure}[htbp]
    \includegraphics[width=9.0cm]{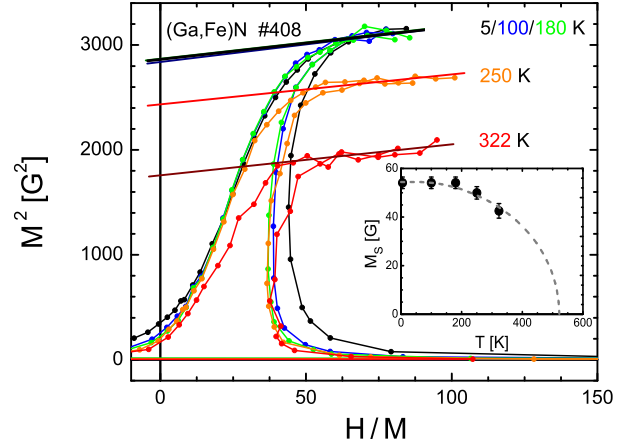}
        \caption{Determination of the spontaneous magnetization at various
temperatures obtained by plotting the square of the magnetization
vs.~the ratio of the magnetic field to the magnetization (Arrot
plot, sample \#408). The inset shows the established spontaneous
magnetization as a function of temperature, whose extrapolation
leads to the apparent Curie temperature of 500~K for the same
sample.} \label{Figure:SQ_Arrot_1}
\end{figure}

\begin{figure}[htbp]
    \includegraphics[width=9.0cm]{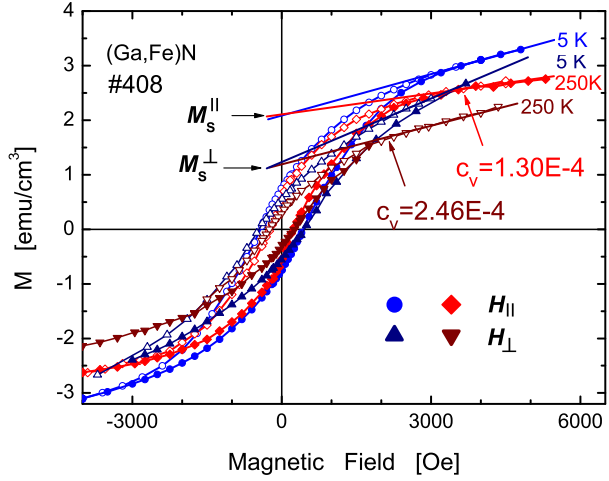}
        \caption{Magnetization measurements for sample \#408 (solid
symbols), indicating that the saturation value of magnetization is
greater for the magnetic field $H_{\parallel}$ along the surface of
the sample (perpendicular to the $c$-axis). Data obtained by
numerical reflection, if only a half of the hysteresis cycle has
been measured, are shown by open symbols.}
    \label{Figure:SQ_408_anisotropy}
\end{figure}

According to the findings summarized in
Fig.~\ref{Figure:SQUID_high_x_200}, the ferromagnetic signal does
not diminish remarkably when the temperature is increased to 200~K,
and it persists up to above room temperature, as shown in the inset.
The set of magnetization curves obtained in the full available
temperature and field range allows to establish the temperature
variation of spontaneous magnetization $M_{\mathrm{S}}$, determined
from the Arrot plot, as depicted in Fig.~\ref{Figure:SQ_Arrot_1}. If
the ideal Brillouin $M_{\mathrm{S}}(T)$ dependence is assumed, the
data displayed in the inset point to an apparent Curie temperature
$T_{\mathrm{C}}$ of $\sim$~500~K. The ferromagnetism in question
reveals a well defined anisotropy.
Figure~\ref{Figure:SQ_408_anisotropy} shows that the saturation
value of magnetization is greater for the magnetic field
perpendicular to the wurzite $c$-axis. Interestingly, a similar
anisotropy was observed in the case of
(Zn,Co)O.\cite{Venkatesan:2004_a}

We want to underline here that we do not take all the above
ferromagnetic signatures as a solid proof of a uniform ferromagnetic
state of (Ga,Fe)N. In the material under consideration here, the
average concentration of magnetic ions is far below the percolation
limit for the nearest-neighbor coupling, and at the same time the
free-carrier density is too low to mediate an efficient long-range
exchange interaction. Hence, we take the presence of high
temperature ferromagnetism as an indication of the non-uniform
distribution of the Fe ions over the GaN lattice already
demonstrated $\textit{via}$ TEM analysis of the structures.

\begin{figure}[htbp]
    \includegraphics[width=9.0cm]{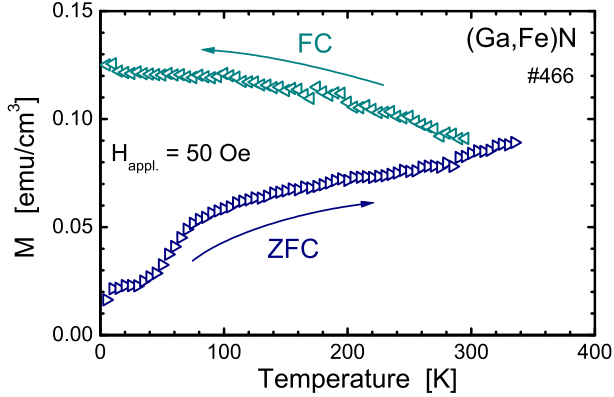}
        \caption{Magnetization of sample \#466 measured at
$H_{\mbox{\tiny{\emph{appl.}}}}$~= 50~Oe after cooling the sample
in absence of the external field (ZFC) and during cooling in
the same $H_{\mbox{\tiny{\emph{appl.}}}}$. Qualitatively the same
results are obtained at $H_{\mbox{\tiny{\emph{appl.}}}}$~=
200~Oe.} \label{Figure:SQ_FC_ZFC_1}
\end{figure}

In order to shed some light on the origin of the ferromagnetic
behavior presented in the last figures, we discuss at first the
results from zero-field cooled (ZFC) and field cooled (FC)
magnetization measurements at a low magnetic field of 50~Oe, given in
Fig.~\ref{Figure:SQ_FC_ZFC_1}. The markedly different behavior,
namely the increasing FC-ZFC difference on lowering temperature, is
indicative of the fact that the sample is in a magnetically frozen
state having its response determined by thermally activated
processes across some energy barriers. We further note that the ZFC
and the FC magnetizations are characterized by temperature gradients
of opposite sign without a maximum appearing in the ZFC curve. Such
a behavior can be expected for an ensemble of magnetically
anisotropic ferromagnetic particles thermally cycled below its
blocking temperature ($T_{\mathrm{B}}$). At $T < T_{\mathrm{B}}$ the
moments of some of the particles are blocked and unable to surmount
the magnetic energy barriers at the time scale of the experiment.
The energy barriers in the vicinity of $H=0$, cannot be overcome
without an increase in the thermal energy (or an external magnetic
field). For a system of non-interacting, uniaxial particles the
energy barrier for the magnetization reversal is $\Delta E = KV$,
where \emph{V} is the particle volume and \emph{K} is the anisotropy
energy density. Therefore, the actual magnetic behavior depends on
the measuring time ($t_m$) of the specific experimental technique
with respect to the time $t$ required to overcome the energy barrier
(for DC magnetic studies a $t_m$ of 100~s is usually adopted).
Accordingly, for a system of identical non-interacting uniaxial
particles, the following simple condition for the evaluation of
$T_{\mathrm{B}}$ is employed:
\begin{equation}\label{Eq:TB}
T_{\mathrm{B}} \simeq KV/25k_{\mathrm{B}}.
\end{equation}
Therefore, $T > T_{\mathrm{B}}$ defines the onset of
superparamagnetic behavior for this system, whereas below
$T_{\mathrm{B}}$ an irreversible behavior is expected. Within this
line of interpretation, our experiment
(Figs.~\ref{fig:SQUID_low_ferrom}-\ref{Figure:SQ_FC_ZFC_1}) points
then to a large, at least above room temperature $T_{\mathrm{B}}$
for (Ga,Fe)N layers.

\begin{figure}[htbp]
	\includegraphics[width=8.6cm]{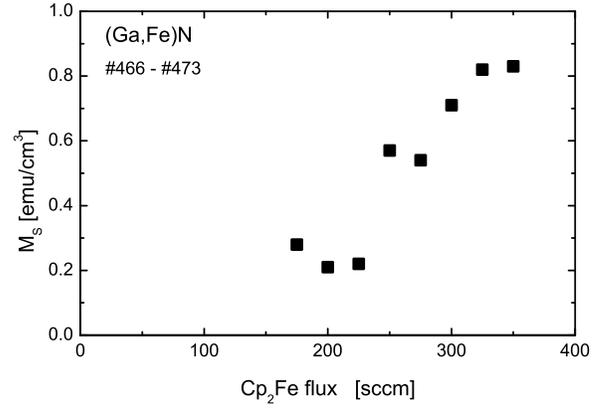}
		\caption{Saturation magnetization of the ferromagnetic component
extracted from \emph{M(H)} curves at 200~K for the \#466 - \#473
layer series versus the Fe precursor flow rate.}
	\label{Figure:Ms_vs_fluxFe}
\end{figure}

This is a rather high value and it indicates that the ferromagnetic
particles in (Ga,Fe)N contributing to the ferromagnetic response
must be characterized by substantial values of the $KV$ product. The
existence of fine entities in the studied samples is corroborated by
our TEM analysis, revealing the presence of wurzite nanocrystals
embedded in the host GaN. The substantial amount of these
nanocrystals, as measured by the value of $M_{\mathrm{S}}$, is
clearly triggered by higher flow rates (above $\sim$150~sccm) -- see
Fig.~\ref{Figure:Ms_vs_fluxFe} -- and undoubtedly correlates with
the nominal Fe concentration in the layers. Furthermore, as already
pointed out, EDS results demonstrate that the Fe concentration is
largely enhanced within the nanocrystals.

All these findings imply that in the case of Fe contents above the
solubility limit (Ga,Fe)N undergoes a spinodal decomposition into
regions with small and large concentrations of magnetic ions, so
that most of Fe ions resides in the magnetic nanocrystals. Now,
taking the volume of these Fe-rich nanocrystals as $V=1000$~nm$^3$
on average, we end up with a requirement for \emph{K} to be $\simeq
20 \times 10^5$~erg/cm$^3$ to reproduce a $T_{\mathrm{B}}$ of about
500~K, as the magnetization studies indicate. This value is four
times higher than that of bulk cubic iron and it is a half of that
of hexagonal cobalt. However, the estimated large value for \emph{K}
should be treated rather as an upper limit, since $T_{\mathrm{B}}$
can get enlarged above the value predicted by Eq.~\ref{Eq:TB} if, as
a result of the magnetic interaction between particles, an ordering
of the magnetic moments takes place. Moreover, we note that the
nanocrystals are not monodispersed and that the distribution in
their volumes gives rise to a distribution in the blocking
temperatures. Thus, Eq.~\ref{Eq:TB} gives the mean
$T_{\mathrm{B}}$ for the system, so that irreversible effects occur
also above $T_{\mathrm{B}}$ and, therefore, the requirements for the
magnitude of $K$ are considerably relaxed.

\section{Summary and outlook}

Our results on the GaN:Fe and (Ga,Fe)N systems have reemphasized the
importance of two related and generic aspects of magnetically doped
wide band-gap semiconductors. First, most of the substitutional
magnetic impurities give rise to the presence of deep levels
originating from open $\textit{d}$ or $\textit{f}$ shells of the
impurity atoms. The presence of such levels strongly affects the
electrical and the optical properties of these materials systems. At
the same time, co-doping with shallow impurities, by changing the
charge and, thus, the spin state of the magnetic ions, can serve to
alter the magnetic response.  Second, though depending on the growth conditions, the TM solubility limit
is rather low and typically --
except for Mn -- it is hard to introduce more than 1\% of magnetic
impurities into randomly distributed substitutional sites. Due to
the low density and in absence of band carriers, the impurity spins can
be regarded as decoupled. Accordingly, pertinent properties of the
system can be adequately described by the single-impurity
theory.\cite{Zunger:1986_a} The series of transport, optical, EPR,
and magnetic experiments that we have reported here for MOCVD grown
(Ga,Fe)N and for (Ga,Fe)N co-doped with either Si or Mg has
substantiated the notion that the material characteristics are
determined by the Fe$^{3+}$/Fe$^{2+}$ acceptor-like deep level.
Thus, the relative concentrations of Fe$^{3+}$ and Fe$^{2+}$ ions
and, hence, the relative importance of the Curie and Van Vleck
paramagnetism is defined by the density of shallow acceptors and
donors. This means, in particular, that if the acceptor
concentration were high enough to give rise to the presence of a
sufficiently increased density of weakly localized or delocalized
valence band holes, the carrier-mediated coupling between spins
localized on Fe$^{3+}$ ions would operate.\cite{Graf:2003_a} Our
experiments have indicated also that the solubility limit is about
0.4\% in the case of Fe in GaN under our growth conditions.

A number of scenarios can be realized when increasing the Fe
concentration beyond the solubility limit. One would be the
appearance of Fe in interstitial positions, where it will act as a
double donor. We have indeed found a strong increase in the electron
concentration in this regime. EXAFS or Rutherford backscattering experiments might
allow to detect the presence of interstitial Fe.  A further scenario
is represented by the crystallographic phase separation in the form
of precipitates of Fe or Fe-compounds. Finally, an appealing
possibility is the existence of spinodal decomposition into regions
with respectively high and low concentrations of the magnetic
constituent. Our TEM results have demonstrated the presence of
coherent Fe-rich nanocrystals embedded and stabilized by the GaN
host. Furthermore, the TEM imaging shows that these nanocrystals are
not distributed uniformly over the layer but float towards the
surface during the growth process. The resulting inhomogeneity in
the layer structure along the growth direction may account for the
increased surface roughness at high doping level, as observed by
AFM. This inhomogeneity may also give reason for the simultaneous
presence of deep Fe acceptors and conducting electrons demonstrated
by our transport, EPR, and SQUID experiments. We argue that the
Fe-rich nanocrystals account for the high-temperature ferromagnetic
properties of the (Ga,Fe)N films grown by MOCVD and MBE as well as
of GaN implanted with Fe. We note that the GaN host may stabilize
nanocrystals of a composition and structure non-existing in the case
of a free standing Ga-Fe-N material systems. Actually, such a
situation has lately been found in the case of MBE grown (Ge,Mn),
where spinodal decomposition into Ge and novel Mn-rich (Ge,Mn)
nanowires has been revealed.\cite{Jamet:2006_a} Furthermore, it has
recently been suggested by one of us\cite{Dietl:2006_a} that the
aggregation of magnetic nanocrystals can be modified by altering the
charge state of magnetic impurities during the growth or by thermal
treatment during or upon deposition. This can obviously be
accomplished by co-doping with shallow donors and acceptors.
Experiments in this direction are on the way. While our paper is
concerned with one material system, namely GaN:Fe -- (Ga,Fe)N, there is ground
to conjecture that spinodal decomposition and inhomogeneity along
the growth direction revealed by our work, as well their influence
on transport, optical, and magnetic characteristics, are generic
properties of diluted magnetic semiconductors and diluted magnetic
oxides showing high apparent Curie temperatures. Within this model,
the spontaneous magnetization originates from magnetic moments of
ferromagnetic or ferrimagnetic nanoclusters or from uncompensated
spins at the surface of antiferromagnetic nanoparticles. We are
confident that the identification of the actual chemical nature of
nanocrystals in particular matrices and the determination of
the mechanisms accounting for the elevated blocking
temperatures necessary to make the ferromagnetic-like characteristics to
survive up to high temperatures, will attract considerable attention
in the years to come.

\begin{acknowledgments}
 This work was partly supported by the
 Austrian Fonds zur F\"orderung der wissenschaftlichen Forschung - FWF
 (projects FWF-P17169-N08 and FWF-N107-NAN)
 and by the EC project NANOSPIN (FP6-2002-IST-015728). The set-up
 for $\textit{in-situ}$ XRD has been provided by
 PANalytical B.V., Almelo, The Netherlands.
 We are grateful to O. Fuchs for his precious technical support.
\end{acknowledgments}

\section*{Appendix: Experimental difficulties and precautions in SQUID measurements}

\begin{figure}[htbp]
  \centering
       \includegraphics[width=9.0cm]{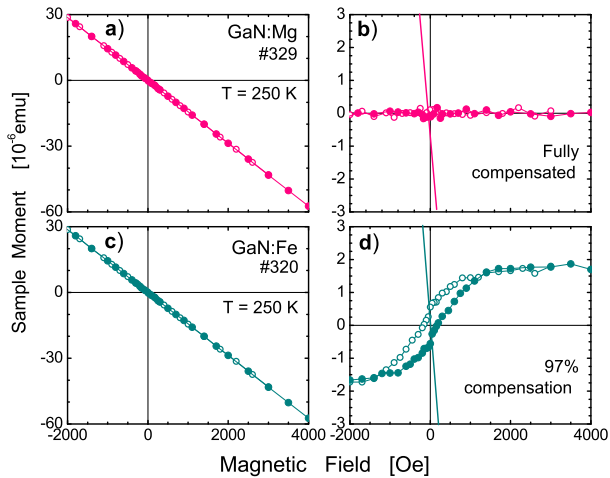}
    \caption{Typical examples of magnetic moment vs.~magnetic field $m(H)$
recorded for the layers presented in this paper. Left panels: raw
data. Right panels: after subtracting the major linear component. The
high slope lines on the left panels indicate the uncompensated
$m(H)$ data. Data obtained by numerical reflection, if only a half
of the hysteresis cycle has been measured, are shown by open
symbols.}
   \label{Figure:SQ_Compensation.eps}
\end{figure}

$\textit{General remarks}$ -- the (Ga,Fe)N layers are grown on
330~$\mu$m thick sapphire substrates, so they constitute only a fraction
of the whole volume of the investigated material, and their
magnetic moment is very-small to small when compared to the
diamagnetic signal of the substrate. Since it is unfeasible to
remove the sapphire substrate, the
magnetic moment of the (Ga,Fe)N system has to be established by
post-measurement subtraction of the known magnetic
moment of a Fe-free, but otherwise identical GaN/sapphire reference
sample. This in general does not pose any experimental difficulties
if the signal to be investigated is at least comparable to the
substrate signal. In the case of our epitaxial layers of
magnetically diluted materials, the weak magnetic response, as shown
later, requires typically more than 90\% compensation, approaching
99\% for the lowest Fe concentration. Furthermore, the overall
measured signals are rather weak, so there is little room left for a
significant improvement of the absolute accuracy of the experimental
data after the compensation, as we are approaching the base noise
level of the magnetometer set up. Results of typical measurements
are depicted in Fig.~\ref{Figure:SQ_Compensation.eps}, exemplifying
the necessity for high quality, distortion free measurements.
Additionally, any experimental flaws, short-cut procedures or
negligences in the sample handling have severe adverse effects on
the reliability of the final results, as they get strongly
magnified by the high degree of compensation analysis required to
determine the signal from these layers. To preserve the credibility
of the measurement, tight constrains on the whole measurement
process are imposed and can be generalized as: contamination-free handling of the
samples, magnetically-clean mounting in the SQUID magnetometer,
and credibility in the determination of the magnetic moment by
the magnetometer software.

\paragraph{Handling of the samples}
We perform all the manual operations, including the sample mounting
on the magnetometer sample holder, in a laminar flow cabinet
equipped with an absolute filter. Also,  we avoid to bring the
samples into contact with any ferrous materials, including metal
tweezers,\cite{Abraham:2005_a} since (stainless) steel tweezers are
a well recognized source of contamination\cite{Koo:1995_a} of,
$\textit{e.g.}$, silicon wafers and their magnetic signature can be
comparable with the magnetic response expected in our
studies.\cite{Abraham:2005_a} However, contrary to Si, where the
contamination can occur either by contact chemical reaction or by
mechanical scratching, we believe that in our case mostly a
mechanical grinding of the softer metal by the much harder
sapphire/GaN is the major source of magnetic contamination.
Therefore, we employ only titanium-/plastic-based tweezers. Another
potential source of specimen contamination is related to the dicing
of the sapphire wafer, as it requires a diamond saw. For the
preparation of our samples, we employ only plastic-based cutting
wheels on the wafer coated with photoresist for additional
protection. After dicing, the samples are rinsed with organic
solvents in an ultrasonic cleaner. We verified that none of these
substances carries any detectable amounts of ferrous metals.

\paragraph{Sample holder and sample mounting}

In the present study we utilize a home-made SQUID magnetometer built
into a standard cryostat operating from 5 to 325~K in continuous
flow mode. Although equipped with a 1~T superconducting coil, the
set-up has been optimized for very weak field operation and the
noise level, increasing with the field, limits our experimental
abilities to $\sim 0.5$~T. We use a commercially available SQUID
sensor and the matched electronics, connected to a computer based
acquisition system running under an home developed code. The sensing
end of the set-up, like most modern similar systems, utilizes the
controlled tripping of the sample (with the sample holder) through
sensing coils arranged in the 2nd order gradiometer configuration.
Ideally, only the sample to be measured produces the signal. Its
position-related voltages generated by the SQUID electronics during
this movement form the base for the  numerical procedures necessary
for extraction of the magnetic moment. In general, a gradiometer
configuration allows the use of any uniform extended object
-- rod-like shaped -- as a sample holder, since if its length is much
larger than the extent of the pick-up coils, the sample holder
produces no output according to symmetry considerations. In reality,
the limited length of the sample chamber does not allow
sufficiently long sample holders, resulting in a non-linear
position-dependent voltage disturbing the signal from the sample,
particularly in the most sensitive ranges. This parasitic
contribution can be largely reduced by
employing magnetically weak substances, preferably diamagnets,
shaped in the form of small tubes or rods. Nothing, however, will
compensate for the shape variations and for the  distribution of
inhomogeneous magnetic impurity inside the rod material. The former
result from all sort of mechanical imperfections (e.g. variable
cross section area, scratches, etc.), the latter may be an inherent
property of the chosen material. Both generate position-dependent
voltages that modulate or distort the signal from the sample in an
unknown and a-priori unpredictable $T$- and $H$- dependent way,
corrupting, in this way, the results of the measurements in an
unrecoverable way. Therefore, the testing of the sample holder and
the selection against the parasitic magnetic signal turns out to be
of a great importance for this study.

Currently, the most popular choice for magnetometer sample holders
are drinking straws, not only because of their low magnetic
signature, but also due to the flexibility allowed in a glue-free
mounting of samples with dimensions comparable with the diameter of
the straw. For the purposes of our studies, we have undertaken an accurate
testing of the available materials, focusing in particular on those
directly provided by Quantum Design.\cite{QD} Some 30 straws were
scanned along their length at 3000, 1000~Oe, and at remanence,
usually at low $T$ (10~K), and occasionally at 200~K. All tested
straws have been thoroughly cleaned in organic solvent prior to the
tests. The collected position dependent signals are processed
according to the same numerical steps employed to establish
the magnetic moment in the magnetometer, so that the spatial
distribution of the magnetic moment along the straws could be
established. We have found that although on average the straws have
a small magnetic signature suitable for general magnetic
investigations (exhibiting moment variations below
$2\times$10$^{-7}$ emu on a one inch span near their central part),
only two went below $\sim5\times$10$^{-8}$ emu, reaching in this way
the upper limit required for our studies. Instead of using
potentially problematic straws, we opted for $\sim$~20~cm long and 1
or 2~mm wide Si strips cut from 8" commercial wafers.\cite{WVRoy}
After a thorough testing, we have found that all the strips conform
to our stringent requirements concerning the magnetic uniformity.
Additionally, the Si strips ensure an excellent rigidity and allow
for an easy sample mounting in various in-plane orientations. The
major drawback of switching to Si strips is the necessity of using a
glue to fix the sample onto the strip. After testing many substances
for a possible contribution to the measurement as well as for
mechanical reliability and low temperature behavior we settled for
the heavily diluted GE varnish, which best fulfills all
requirements. We found that only the recently available orange
version,\cite{GE_varnish} as opposed to the dark red one,
shows no magnetic signature down to 5~K and up to 3000~Oe even when
examining $\sim$10 times the amount normally needed for mounting. It
must be noted, that the usage of narrow strips may lead to larger
sample off centering than in the case of wider straws. In our case
however, owing to the relatively large diameter of our pick-up coils
we determined this effect to be below 0.5\%,\cite{Zieba:1993_a} i.e.
within the typical run-to-run non-reproducibility of our
system.

\paragraph{Reduction of the data}

Typically the investigated magnetic moment is established by a
numerical fitting (least square regression) of an ideal, point
dipole response function of the gradiometer pickup coils into the
set of voltages collected during the tripping of the sample through
the coils. In the case of macroscopically large samples of rectangular
shape the form of the real
response function starts to differ noticeably if the specimens
occupy more than a few percent of the volume of the pickup
coils.\cite{Stamenov:2006_a} As a consequence, even the best
possible fitting routine will not establish the accurate sample
moment, unless the response function gets adequately modified. It
is relatively easy to extend the point dipole response function to
account for uniformly magnetized body of a given
length\cite{Zieba:1993_a,Blott:1993_a,Stamenov:2006_a} and this
length parameterized function is routinely used in our system to
extract a length-independent magnetic moment.
This however seems to be the only one feasible to compensate
sample size effects. The account of other detrimental effects like
non uniform sample extension in the plane perpendicular to the
magnetometer axis or displacement of the whole sample aside from
the axis is too complex to get generalized in any sort of
universally parameterized function. So, in order to minimize their
adverse influence on the experimental results, we confined
ourselves to one sample size for magnetic measurement and
additional care is taken to mount the samples on Si sticks in a reproducible way.
For completeness, we note that from the results of
numerical modelling\cite{Zieba:1993_a,Stamenov:2006_a} a maximum
1\% of moment distortion is expected for the samples used in the
experiment. Therefore, we can safely conclude that although
potentially large, the sample-size-related effects should not
systematically alter the magnetometer results before and after the
compensation.

\bibliographystyle{apsrev}

\end{document}